\documentclass[preprint, amsmath,amssymb,aps,pra]{revtex4-2}

\usepackage{natbib}
\setcitestyle{sort&compress,numbers}
\usepackage[switch]{lineno} 
\usepackage{amsfonts}
\usepackage{graphicx}
\usepackage{epstopdf}
\usepackage{algorithmic}
\usepackage{graphicx,color}
\usepackage[section]{placeins}
\usepackage{subfig}
\usepackage{epsfig}
\usepackage{amsmath}
\usepackage{hyperref}
\usepackage{cleveref}
\crefformat{figure}{#2Fig.~#1#3}
\crefmultiformat{figure}{#2Figs.~#1#3}{ and~#2#1#3}{, #2#1#3}{ and~#2#1#3}

\crefrangeformat{figure}{#3Figs.~#1#4 to~#5#2#6}

\crefformat{appendix}{#2Appendix~#1#3}
\crefformat{enumi}{#2Step~#1#3}
\crefmultiformat{enumi}{#2Steps~#1#3}{ and~#2#1#3}{, #2#1#3}{ and~#2#1#3}

\crefformat{section}{#2\S~#1#3}
\crefmultiformat{section}{#2\S\S~#1#3}{ and~#2#1#3}{, #2#1#3}{ and~#2#1#3}

\crefformat{equation}{#2Eq.~(#1)#3}
\crefmultiformat{equation}{#2Eqs.~(#1)#3}{ and~#2(#1)#3}{, #2(#1)#3}{ and~#2(#1)#3}

\crefformat{table}{#2Table~#1#3}

\usepackage{verbatim}
\usepackage{chngcntr}
\usepackage{comment}

\begin{document}

\title[Network-based membrane filters]{Network-based membrane filters: Influence of network and pore size variability on filtration performance}
\thanks{This work was supported by NSF Grants No. DMS-1615719 and No. DMS-2133255.}

\author{Binan Gu}
\email{Corresponding emails: bg263@njit.edu, kondic@njit.edu and linda.cummings@njit.edu}
\author{Lou Kondic}%
\author{Linda J. Cummings}
 
\affiliation{ 
Department of Mathematical Sciences,\\ Center for Applied Mathematics and Statistics,\\
New Jersey Institute of Technology,\\
Newark, NJ 07102, USA
}

\begin{abstract}
We model porous membrane filters as networks of connected cylindrical pores via a random network generation protocol, and their initial pore radii via a uniform distribution of widths that vary about some mean value. We investigate the influence of network and pore size (radius) variations on the performance of membrane filters that undergo adsorptive fouling. We find that membrane porosity variations, independently of whether induced by variations of the pore radii or of the random pore network, are an important factor determining membrane filter performance. Network and pore size variations still play a role, in particular if pore radii variations are significant.  To quantify the influence of these variations, we compare the performance metrics of networks built from pores of variable radii to their (equal porosity) counterparts built from pores of uniform radius. We show that the effect of pore radii variations is to increase throughput, but also to reduce foulant control. 
\end{abstract}

\maketitle

\section{Introduction}

Membrane filtration is an important separation process used in many industrial and commercial applications such as treatment of radioactive sludge, water purification, beer clarification~\citep{Lipnizki2015}, semiconductor and microelectronics processing~\citep{HUANG2011203}, air filtration~\citep{LIU2017375} and membrane bioreactors~\citep{Dizge2011}. Membrane filters used in these applications have a wide range of architectures, ranging from single layer thin porous films to multilayered porous membranes~\citep{CHANG2004117,multi_onur_2017} to large scale continuous sheets of layered fibrous material~\citep{hepa,hepa2,hepa3}. 

Many models to describe the underlying membrane pore structure and/or geometry have been proposed and studied in recent years. For example, there are simple theoretical models to analyze the performance of membranes composed of multiple layers of different porous materials~\citep{fong_multi,multi_onur_2017}, or membranes with simple branched structures that can incorporate porosity gradients (\citep{sanaei2018flow,POLYAKOV200881}, among many others). As recent advances in imaging techniques have greatly contributed to the ability to compare such structural models to experiments (\citep{martinez,SUN200793,BLUNT2013197,3d_image}; also see~\citep{imperial} for detailed images), more sophisticated models of membrane architecture have also been formulated, with a recent focus on accurate modeling of membrane filters with a network-type structure, for example, membranes where the solid component is comprised of fibres (so-called node-fibril type membranes) or those that transport feed through networks of capillaries~\citep{krupp_2017,griffith_jms,gu_network_2021} (also used in the context of condensate banking~\citep{reis}). 

The present paper focuses on the latter type of pore network model, and our analysis of pore size variations builds on a model introduced previously in~\citep{gu_network_2021}. Briefly, a network model involves vertices and edges that represent pore junctions and throats respectively (referred to simply as {\it junctions} and {\it pores} from hereon respectively). 
Each pore is assumed to be a circular cylinder of fixed radius, with Hagen-Poiseuille flow, and conservation of fluid flux is imposed at each junction. Foulant is advected through pores by fluid and deposits on pore walls via adsorptive fouling.  Membrane filter performance is analyzed by recording total throughput (volume of filtrate collected during the lifetime of the filter) and accumulated foulant concentration at the membrane outlet. 

The primary findings of~\citep{gu_network_2021} that motivate the present work are the following relationships. Firstly, the initial porosity (pore void volume divided by total domain volume) of the network is demonstrated to be an important material feature that predicts total throughput by a power law, which holds particularly well when initial porosity $>0.5$. Secondly, it was found that the accumulated foulant concentration at the membrane outlet decays exponentially with the initial tortuosity of the network (defined as the average distance travelled by a fluid particle from membrane inlet to outlet, see \cref{app:tor} for a detailed definition). 


In addition to the features discussed above, pore size (radius) variation is another important aspect of membrane filter design, which has been investigated by a number of authors, in particular regarding how it affects membrane selectivity and particle retention in a variety of applications~\citep{zydney_93,zydney_95,zydney_10,LIU2021119097,TAYLOR2021119436}. It has not, however, been extensively studied in the context of network models for membrane filters: though such models allow for cylindrical pores with a distribution of lengths \citep{griffith_jms,gu_network_2021}, random variations of the pore radii have not, to our knowledge, been considered. Accordingly, in this paper, we focus on membrane filters whose pores may be considered as a network of interconnected capillaries of different initial radii. Henceforth in this work, the phrase ``pore size variations'' refers to variations in the initial pore radii, unless otherwise specified.

We present a novel characterisation of the membrane pore network and model pore size variations via a set of random initial conditions for the radii of pores. Our assumption on the pore size distribution, in contrast to the log-normal distribution usually assumed in applications~\citep{ZYDNEY1994293}, has the advantage of incurring only one additional parameter -- noise amplitude, as opposed to a mean and variance in a log-normal distribution approach -- while still capable of capturing effects such as inhomogeneities during the manufacturing process. We use our network representation to investigate the effect of pore size variations on the performance of membrane pore networks, as characterized by foulant concentration and total throughput of filtrate over the filter lifetime. 

The paper is structured as follows:
in \cref{sec:setup}, we describe the network model setup and introduce the key performance metrics. In \cref{sec:methods}, we set out our investigation strategy for pore size variation in an algorithm and declare the main nomenclature used in the analysis. In \cref{sec:results}, we present and discuss our main results and in \cref{sec:conclusion}, we conclude our findings. 

\section{Setup: General Pore Networks}\label{sec:setup}
In this section we construct our model of a membrane filter represented by a random network of connected cylindrical pores. In \cref{sec:network}, we describe how we generate a network that represents the internal pore structure of a membrane filter; in \cref{sec:fluid,sec:fouling}, we outline the governing equations for the Hagen-Poiseuille fluid flow, for the advection of foulant particles carried by the flow, and for the pore evolution in time; in \cref{sec:notations} we specify our notation; in \cref{sec:scales}, we provide the relevant scales; and in \cref{sec:metrics}, we introduce two metrics to characterize the performance of a membrane filter with a specified pore network.

\subsection{Network Generation\label{sec:network}}
\begin{figure}
    \centering
    \includegraphics[scale = 0.4]{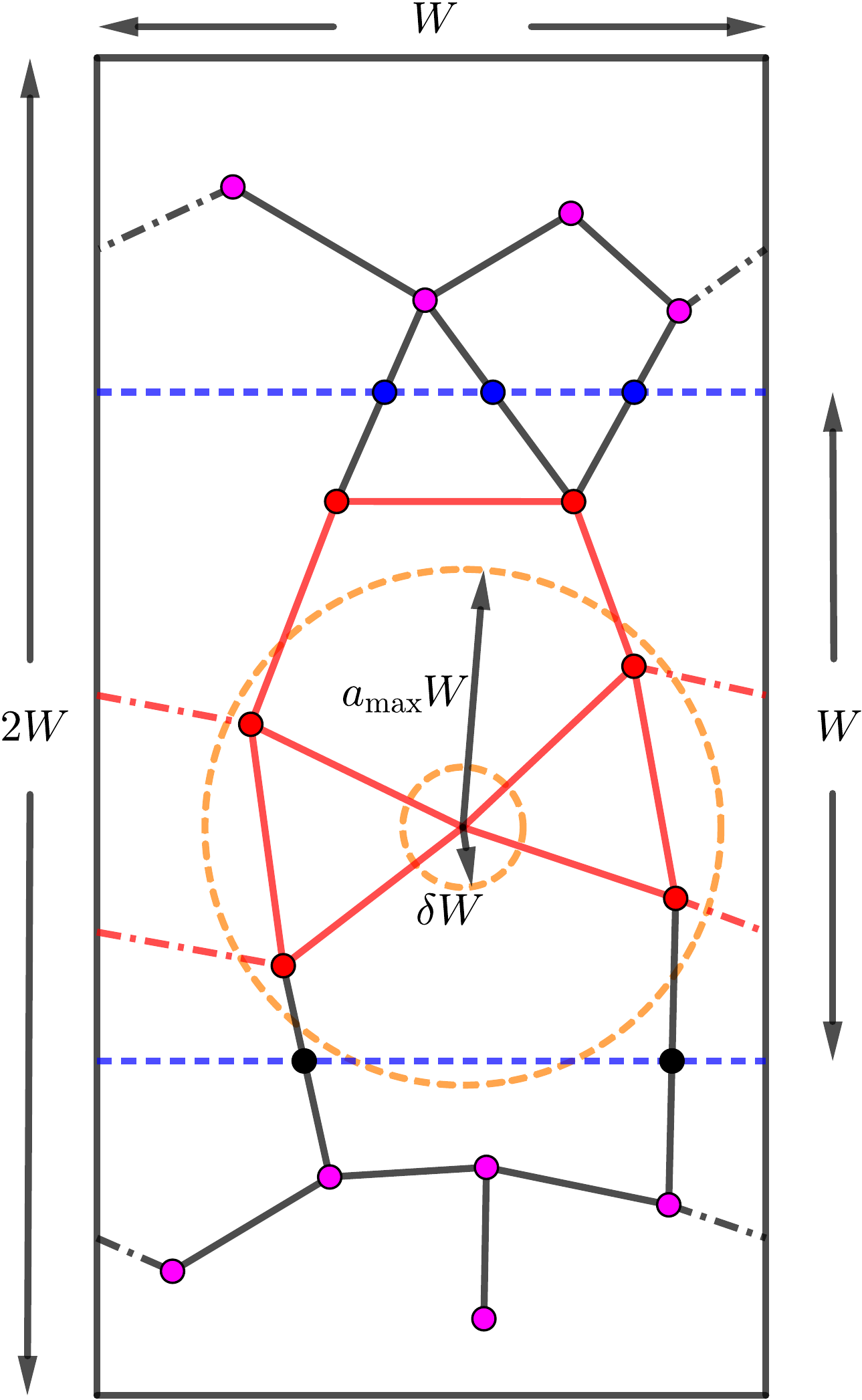}
    \caption{2D schematic of the 3D network generation with periodic boundary conditions showing: interior junctions $\mathcal{V}_{\rm int}$ ({\color{red}red filled circles}); pore inlets $\mathcal{V}_{\rm in}$ ({\color{blue}blue filled circles}) and outlets $\mathcal{V}_{\rm out}$ (black circles) induced by the cutting process; the cutting planes {\color{blue}blue dashed lines}; discarded points ({\color{magenta}magenta filled circles}). Solid lines represent pores, while dash-dotted lines are pores that arise from the periodic boundary condition ({\color{red}red} are interior to the membrane). $a_{\rm max} W$ and $\delta W$ are prescribed maximum and minimum pore lengths respectively.}
    \label{fig:3d_bdy_gen}
\end{figure}

\begin{figure}
    \centering
    \includegraphics[scale=0.5]{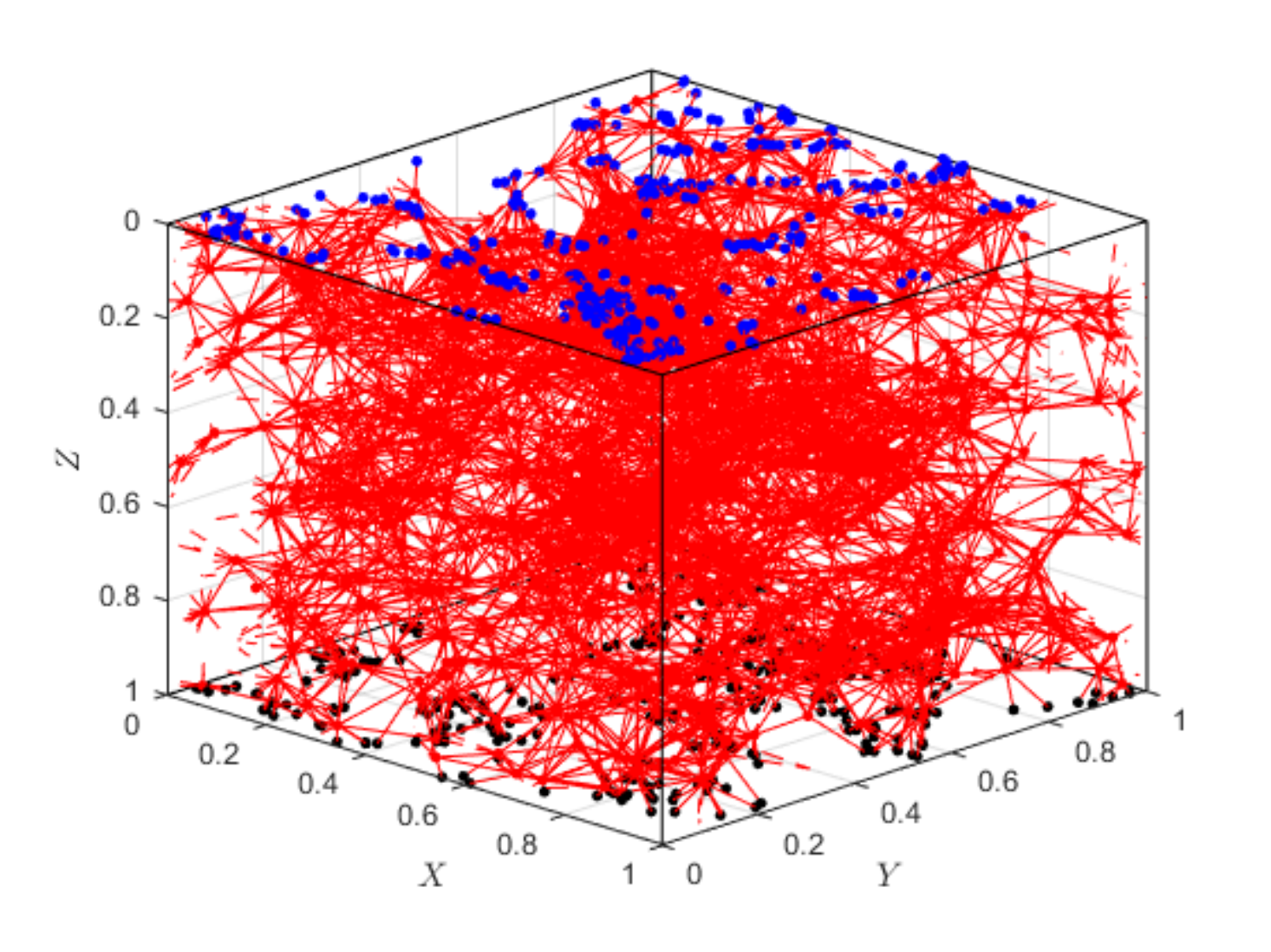}
    \caption{Schematic of a 3D network with $a_{\rm max}=0.15$ and $N_{\rm total} = 2000$. {\color{red}Solid red lines} are interior pores; {\color{red} dashed red lines} are pores created by the periodic boundary conditions. {\color{blue}Blue dots} are inlets. Black dots are outlets. }
    \label{fig:3d_realization}
\end{figure}

We first demonstrate the 3D random network generation protocol employed in this work (following~\citep{gu_network_2021}). In \cref{fig:3d_bdy_gen}, for illustration purposes, we show a 2D schematic of this protocol: the domain is a rectangular cell (square prism in 3D) of lateral height two times its horizontal width $W$, assumed to repeat periodically at the lateral boundaries (more details in \cref{app:gg}). We generate a pore network by first uniformly distributing $N_{\rm total}$ (a large integer to be prescribed) points as the {\it junctions} (nodes of the network) inside the domain. We construct {\it pores} (the {\it edges} of the network) by connecting all pairs of points that lie within a prescribed maximal distance $a_{\rm max} W$, but at least some minimal distance $\delta W$ apart. We then cut the prism horizontally at two locations (see {\color{blue}blue} dotted lines in \cref{fig:3d_bdy_gen}), and discard the end pieces to produce a square (or cube in 3D), which will represent an element of a membrane filter. The intersections of the two cutting planes and the pores (edges) naturally form  membrane inlets and outlets, respectively (see~\cref{fig:3d_realization} for a 3D realization, with lengths scaled by $W$).

\subsection{Fluid Flow\label{sec:fluid}}

We next characterize fluid flow in the membrane network. The flow through the pores (which, within the current model, are cylinders of circular cross section) is assumed to obey the Hagen-Poiseuille model, valid provided the pores have sufficiently small aspect ratio (ensured in practice by choice of $\delta$). The Hagen-Poiseuille equation states that fluid flux $Q$ is proportional to pressure difference $\Delta P$ along each pore,
\begin{align}\label{intro_hp} 
    Q&=K\Delta P, \quad 
    K=\frac{\pi R^{4}}{8\mu A},
\end{align}
where the conductance $K$ depends on $R$ and $A$, the radius and length respectively of the pore, and $\mu$ is fluid viscosity. To drive the flow, we prescribe the transmembrane pressure by setting the values on the top ($P_0$) and bottom ($0$) membrane surfaces, while the pressures at interior junctions are unknown. At each junction, we impose conservation of fluid flux, which leads to a system of equations (a discrete Laplace equation for these junction pressures, see \cref{app:gl} for details). Once the pressures are found, flux through each throat is determined by \cref{intro_hp}.

\subsection{Advection and Adsorptive Fouling\label{sec:fouling}}
Feed solution enters the membrane top surface with a fixed foulant concentration (simply referred to as {\it concentration} from hereon), which serves as a boundary condition. Foulant particles are advected by the fluid while depositing on the pore walls, causing pores to shrink and eventually close up, a process known as adsorptive fouling. For each pore, let $C\left(Y,T\right)$ be the particle concentration at any point $Y$ of the pore at time $T$, where $Y$ is a local coordinate measuring distance along a pore in the direction of fluid flow. The concentration satisfies the steady state advection equation
\begin{align}
Q\frac{\partial C}{\partial Y} &= - \Lambda RC, \quad 0\leq Y \leq A,\label{cont_transport}\\
C\left(0,T\right)&:= C_{\rm up}\left(T\right). \label{interior_bc}
\end{align}
Here, the right hand side of \cref{cont_transport} is a sink of concentration that models foulant particles depositing on pore walls as they traverse the edge, with $\Lambda$ a parameter that captures the affinity between foulant particles and membrane material; and $C_{\rm up}\left(T\right)$ in \cref{interior_bc} is the concentration at the upstream entrance of a pore. For pores connected to the membrane top surface (at inlets), we prescribe a constant concentration $C_{\rm up}\left(T\right)=C_{\rm top}$ as the upstream boundary condition that represents the concentration in the feed solution. The foulant concentrations at interior junctions are solved for using conservation of foulant particle flux $QC$; that is, the total combined incoming particle flux from upstream pores must be equal to that entering downstream pores. Similar to the fluid flux conservation, this law can be expressed by a system of equations that accounts for the network connectivity (see \cref{app:gl} and~\citep{gu_network_2021} for more details).

Lastly, as foulant particles deposit on the pore wall, pore radius decreases at a rate depending on local upstream concentration $C_{\rm up}\left(T\right)$,
\begin{equation}
    \frac{dR}{dT}=-\Lambda \alpha C_{\rm up}\left(T\right), \quad R\left(0\right) = R_{0}.
\label{dim-adsorption}
\end{equation}
where $\alpha$ is a parameter related to particle volume (see the appendix of \citep{gu_network_2021} for details). This pore evolution equation closes the model for fluid flow and foulant transport during a membrane filtration process. The filter lifetime $T_{\rm final}$ is reached when the outgoing flux at the membrane's downstream surface falls to zero. 

\cref{dim-adsorption} implicitly assumes that the radius of each pore is a function only of time $T$ and is independent of the local coordinate $Y$ during its evolution, which is determined only by the upstream particle concentration. A more accurate model would be $\partial R (Y,T)/\partial T =-\Lambda\alpha C(Y,T)$; however, the simplification represented by \cref{dim-adsorption} yields only minor differences in model outcome~\citep{gu_network_2021}, while providing significant computational benefit since it 
permits analytical solution of \cref{cont_transport}, leading to immense computational speed-up (more than $100$ times faster than the alternative). In particular, we have verified that 
the more accurate model leads to a modest increase (at most $10\%$) in total throughput. The difference in accumulated foulant concentration is smaller still, and differences in both performance measures are particularly insignificant when $a_{\rm max}$ (the ratio of the maximum pore length to the membrane thickness) is small. The computational savings afforded by this simplification facilitate the large number of simulations carried out in \cref{sec:methods}, needed to obtain reliable statistics.

\subsection{Network Notations}\label{sec:notations}
In \cref{sec:fluid,sec:fouling}, we introduced fluid flow and adsorptive fouling in each individual pore and described the conservation laws that govern the physical quantities of interest in a pore network. Since describing the dynamics of such quantities requires coupling of all junctions and pores, we introduce the following indexing scheme. We use a single index $\left(\cdot\right)_i$ to indicate junction dependence only, e.g. pressure $P_i$ at junction $i$. We use a double index $\left(\cdot\right)_{ij}$ to indicate dependence on the pore connecting junctions $i$ and $j$, e.g. flux $Q_{ij}$. We summarize all physical quantities in \cref{nomenclature}. 

\begin{table}
\begin{tabular}{|c|c|}
\hline 
Dimensional Quantity at time $T$  & Symbol\tabularnewline
\hline 
\hline 
Length of pore $ij$ & $A_{ij}$\tabularnewline
\hline 
Radius of pore $ij$ & $R_{ij}\left(T\right)$\tabularnewline
\hline 
Membrane unit length & $W$\tabularnewline
\hline 
Maximum pore length & $a_{{\rm max}}W$\tabularnewline
\hline 
Minimum pore length & $\delta W$\tabularnewline
\hline 
Pressure at junction $i$ & $P_{i}\left(T\right)$\tabularnewline
\hline 
Concentration at junction $i$ & $C_{i}\left(T\right)$\tabularnewline
\hline 
Flux in pore $ij$ & $Q_{ij}\left(T\right)$\tabularnewline
\hline 
Deposition Coefficient & $\Lambda$\tabularnewline
\hline 
\end{tabular}
\caption{Key Dimensional Quantities}\label{nomenclature}
\end{table}

\subsection{Scales\label{sec:scales}}
We nondimensionalize the fluid flow and foulant transport model described in the previous sections as follows:

\begin{equation}\label{eq:scaling}
    \begin{gathered}
    A_{ij} = W a_{ij}, \quad \left(R_{ij},R_{ij,0}\right) = W \left(r_{ij},r_{ij,0}\right), \\
    P_i = P_0 p_i, \quad \left(C_i,C_{ij}\right) = C_{\rm top} \left(c_i,c_{ij}\right),\\
    \Lambda = \frac{\pi W P_0}{8\mu}\lambda, \quad T = \frac{W}{\Lambda \alpha C_{\rm top}}t,
    \end{gathered}
\end{equation}
where the upper case symbols are dimensional quantities (listed in \cref{nomenclature}). Parameter values are listed in \cref{parameter_values}. We comment briefly regarding the parameter $\lambda$: its numerical value is expected to be small since such values promote an efficient use of the filter by allowing particles to permeate deeper into the membrane interior. Our choice of the specific value, while arbitrary, satisfies this requirement. Further work, beyond the scope of this paper, would be needed to explore the influence of the exact value given to this quantity.  Provided $\lambda$ is sufficiently small we do not anticipate any significant modifications of the trend of the results that follow, since the main role of this parameter is to determine the time scale of the problem.  

\begin{table}
\begin{tabular}{|c|c|c|}
\hline 
 
Parameter & Symbol & Values\tabularnewline
\hline 
\hline
Initial number of junctions~~ & $N_{\rm total}$ & $500,775$  \\
\hline 
Maximum pore length & $a_{\rm max}$ & $0.3$  \\
\hline 
Minimum pore length & $\delta$ & $0.06$  \\
\hline 
Unperturbed pore radius & $r_0$ & $0.01$  \\
\hline 
Noise Amplitude & $\beta$ & $0.06,0.25,0.5,0.7$  \\
\hline 
Number of networks & $N_{\rm net}$ & $1000$  \\
\hline 
Number of noise realizations & $N_{\rm noise}$ & $500$  \\
\hline 
Deposition coefficient & $\lambda$ & $5\times 10^{-7}$ \\
\hline 
\end{tabular}
\caption{Key parameters}\label{parameter_values}
\end{table}

\subsection{Performance Metrics\label{sec:metrics}}
We evaluate the performance of a membrane network using the following two metrics: 1) total throughput ($H$) and 2) accumulated foulant concentration at membrane outlet ($C$). Total throughput is the total volume of filtrate collected at the membrane outlet over the lifetime of the membrane network filter. Accumulated foulant concentration measures the aggregate concentration of foulant particles in the collected filtrate when the filter is exhausted. Precise definitions of these quantities are given in \cref{app:pm}. 

\section{Investigation Methods}\label{sec:methods}
We study pore size variations by prescribing a random initial condition for each pore radius, $r_{ij,0}$. More precisely, we consider perturbations to uniform pores by imposing a multiplicative noise,
\begin{equation}
    r_{ij,0} = r_0\left(1+\epsilon_{ij}\right),
    \label{eq:perturbation}
\end{equation}
where $\epsilon_{ij} \sim {\rm Unif}\left(-\beta,\beta\right)$ is a uniform random variable with noise amplitude $0<\beta<1$, independent for each pore. For the rest of this work, when we use the terms {\it perturbation} or {\it noise}, we are referring to \cref{eq:perturbation}. 

We will study selected simulation outputs, $F$, as model parameters vary, with a primary focus on the influence of noise amplitude, $\beta$. Specifically, in this work $F$ will be one of the following:
\begin{itemize}
        \item $H$, total throughput (throughput),
        \item $C$, accumulated foulant concentration at membrane outlet (concentration), 
        \item $V$, initial network porosity (porosity), 
        \item $\tau$, initial tortuosity (tortuosity), 
\end{itemize}
where the abridged terms in parentheses are used freely henceforth (see detailed definitions of $H$, $C$ and $\tau$ in \cref{app:pm} and \cref{app:tor}). We also note that since the nondimensional model operates in the domain of a unit cube, initial void volume and membrane porosity are equal in value. We refer to {\it porosity} only hereon.

A principal aim of this work is to compare the influence of two independent sources of randomness, namely, the random network generation process ({\it network variability} henceforth), and the random initial condition for the pore radius that yields pore radius variations ({\it noise variability} henceforth), on statistics of membrane filter performance metrics. Here we describe the methodology of our study, before summarizing the approach as an algorithm with enumerated steps below. 

First, we generate a large number, $N_{\rm net}$, of random membrane networks (per \cref{sec:network}), each with the same initial pore radius $r_0$. We perturb the pore networks via \cref{eq:perturbation} in the following two distinct ways: To probe noise variability, we fix one particular ``typical'' (to be made precise in what follows) network from the $N_{\rm net}$ that were generated, perturb it independently $N_{\rm noise}$ times (for $N_{\rm noise}$ sufficiently large), solve the governing \cref{intro_hp,cont_transport,dim-adsorption} (coupled over the entire network using continuity as described) and collect statistics of performance metrics from these $N_{\rm noise}$ realizations of noise. To probe network variability, we perturb each of the $N_{\rm net}$ networks independently just once, and collect performance statistics from the perturbed networks. 

Perturbing the pore radii inevitably changes the porosity of the network.  This is important, since initial network porosity was shown to influence strongly both of our membrane performance metrics in the unperturbed case \citep{gu_network_2021}. To investigate the importance of such induced porosity changes, we devise the following strategy. First, note that the initial porosity of a network is given by
\begin{equation}
    V = \frac{\pi}{2} \sum_{ij} r_{ij,0}^2 a_{ij} \overset{\cref{eq:perturbation}}{=} \frac{\pi}{2}r_0^2 \sum_{ij} \left(1+2\epsilon_{ij}+\epsilon_{ij}^2\right) a_{ij},
    \label{eq:vol_formula}
\end{equation}
where the last expression implies that perturbed networks have larger porosities on average, since $\epsilon_{ij}^2$ is nonnegative (the factor of 1/2 is needed because the double sum counts every pore twice). For each perturbed network, we obtain its porosity via \cref{eq:vol_formula}. We then impose this porosity on the underlying {\it unperturbed} network by determining a new initial pore radius such that the unperturbed and perturbed networks have the same porosity. This new unperturbed network is referred to as a {\it porosity-corrected} network, from which we also collect performance statistics. With these preparations, we define a `score' as the difference between the outputs of the perturbed and porosity-corrected networks, normalized by the outputs of the porosity-corrected ones. 

Each of the $N_{\rm net}$ membrane networks has a fixed maximum pore length $a_{\rm max}=0.3$, minimum pore length $\delta=0.06$ and initial unperturbed pore radius $r_0 = 0.01$. We vary two parameters: 1) noise amplitude $\beta$; and 2) total number of junctions $N_{\rm total}$. The ensuing study first fixes a porosity by fixing $N_{\rm total}$, while varying $\beta$. All geometric parameters used in the algorithm are listed in \cref{parameter_values}, with their values. 

We summarize the above procedures in the following algorithm:
\begin{enumerate}
    \item ({\it Random network generation}) Choose $N_{\rm total}$. Generate $N_{\rm net}$ unperturbed networks. Compute the initial porosity $V_0$ for each network. \label{algo:step 1}
    \item ({\it Noise perturbation}) Choose noise amplitude $\beta$. \label{algo:step 2}
    \begin{enumerate}
        \item ({\it noise variability}) Fix a typical network and perturb it $N_{\rm noise}$ times independently. Compute the outputs of the perturbed networks, $F_{\rm noise}$, referred to as {\it output under noise realizations}. The typical network is chosen such that its porosity is the closest to the average porosity of the ensemble, a posteriori. \label{algo:step 2a}
        \item ({\it network variability}) Perturb each unperturbed network once independently via \cref{eq:perturbation}. Compute the associated outputs $F_{\rm net}$, referred to as {\it output under network realizations}. \label{algo:step 2b}
    \end{enumerate}
    \item ({\it Porosity correction}) For each perturbed network, consider its underlying {\it unperturbed} equal-porosity network (created by prescribing the appropriate uniform pore radius for its pores; see \cref{app:vc} for a short derivation). This new network is called a {\it porosity-corrected} network, with output labeled $F_{\rm pc}$. \label{algo:step 3}
    \item ({\it Scores}) We construct two scores that characterize the noise and network variability when comparing networks of equal porosity. The two scores are constructed independently from each other, so there is no confusion between notations. \label{algo:step 4}
    \begin{enumerate}
        \item \label{algo:step 4a} ({\it Noise score}) We compute the following score, 
            \begin{equation}
            \widehat{F}_{\rm noise} = \frac{F_{\rm noise} - F_{\rm pc}}{F_{\rm pc}},
            \label{eq:noise_score}
            \end{equation}
            where $F_{\rm noise}$ and $F_{\rm pc}$ are computed via \cref{algo:step 2a,algo:step 3}, respectively. All quantities in \cref{eq:noise_score} are vectors of length $N_{\rm noise}$. The subtraction and division are element-wise. 
        \item \label{algo:step 4b} ({\it Network score}) Perform \cref{algo:step 2b} to obtain the outputs under network realizations $F_{\rm net}$. Compute the following score,
            \begin{equation}
            \widehat{F}_{\rm net} = \frac{F_{\rm net} - F_{\rm pc}}{F_{\rm pc}},
            \label{eq:network_score}
            \end{equation}
            where $F_{\rm pc}$ are computed via \cref{algo:step 3}. All quantities in \cref{eq:network_score} are vectors of length $N_{\rm net}$. The subtraction and division are element-wise.
    \end{enumerate}
    In all discussions below, we refer to the quantities defined by \cref{eq:noise_score} or \cref{eq:network_score} as {\it porosity-corrected} scores.
    \item Obtain the means and standard deviations of model outputs, $\overline{F_{\rm noise}}$ and $\overline{F_{\rm net}}$, and scores, $\overline{\widehat{F}_{\rm noise}}$ and $\overline{\widehat{F}_{\rm net}}$. The means under noise and network realizations are computed by averaging over the number of noise ($N_{\rm noise}$) and network  ($N_{\rm net}$) realizations, respectively. \label{algo:step 5}
    \item Go back to \cref{algo:step 2} with a different $\beta$. \label{algo:step 6}
    \item Go back to \cref{algo:step 1} with a different $N_{\rm total}$ (to vary initial porosity). \label{algo:step 7}
    
\end{enumerate}

All outputs computed in this algorithm are summarized in \cref{outputs}.
\begin{table}
\begin{tabular}{|c|c|}
\hline 
Dimensional Quantity at time $T$  & Symbol\tabularnewline
\hline 
\hline 
Porosity-corrected & $F_{\rm pc}$ \\
\hline 
Noise realizations & $F_{\rm noise}$ ($\overline{F_{\rm noise}}$,$\sigma_{F_{\rm noise}}$) \\
\hline 
Network realizations  & $F_{\rm net}$ ($\overline{F_{\rm net}}$,$\sigma_{F_{\rm net}}$)  \\
\hline 
Noise score & $\widehat{F}_{\rm noise}$ ($\overline{\widehat{F}_{\rm noise}}$,$\sigma_{\widehat{F}_{\rm noise}}$) \\
\hline 
Network score & $\widehat{F}_{\rm net}$ ($\overline{\widehat{F}_{\rm net}}$,$\sigma_{\widehat{F}_{\rm net}}$ ) \\
\hline 
\end{tabular}
\caption{Key outputs. The quantities in parentheses are the mean and standard deviation of the corresponding output.}\label{outputs}
\end{table}

\section{Results and Discussions}\label{sec:results}

In \cref{sec:results_example}, we study the performance metrics under a specific choice of model parameters as an example. First, we compare the noise and network variability of the raw metrics (throughput and concentration) and geometric quantities (porosity and tortuosity). Then, we compare the porosity-corrected scores under noise (per~\cref{eq:noise_score}) and network (per~\cref{eq:network_score}) realizations (steps 4a and 4b in the algorithm, respectively). In \cref{sec:reinforce}, we reinforce the example by a thorough sweep of the parameter space and present our main results. 

\subsection{Detailed Example: Low Porosity Network in Low Noise Regime}
\label{sec:results_example}
In this section, we present a set of results for membrane pore networks in the regime of low noise amplitude perturbations to the pore radii and low initial porosity. We choose this parameter regime as a detailed example because the results at higher porosities are qualitatively similar but more time-consuming to compute. We generate $N_{\rm net}$ networks with an initial number of points $N_{\rm total}$, which yield an {\it ensemble average} initial porosity $V \approx 0.25$ (averaged over $N_{\rm net}$ unperturbed networks). The noise $\epsilon_{ij}$ is realised $N_{\rm noise}$ times for each network, with fixed noise amplitude $\beta = 0.06$ here. We give values of $N_{\rm net},N_{\rm noise}$ and $N_{\rm total}$ in \cref{parameter_values} and have found that these numbers are sufficient to account for the random nature of the network and noise generation protocol; larger numbers produce similar results.

When we study noise variability (per \cref{algo:step 2a} in the algorithm), we fix a typical network with initial porosity very close to the ensemble average, $0.25$.  We first compare the statistics of performance metrics (per \cref{algo:step 4a,algo:step 4b} in the algorithm) and then discuss the similarities and differences in the network and noise scores calculated for the initial porosity and tortuosity of the networks, similarly averaged over many network and noise realizations. In particular, we focus on the two relationships found in \cite{gu_network_2021} as example results: throughput vs. porosity, and concentration vs. tortuosity.

\subsubsection{Throughput and Porosity}

\begin{figure}[!h]
    \centering
    \subfloat[]{\label{fig:3a}\includegraphics[width=.45\textwidth,height=.4\textwidth]{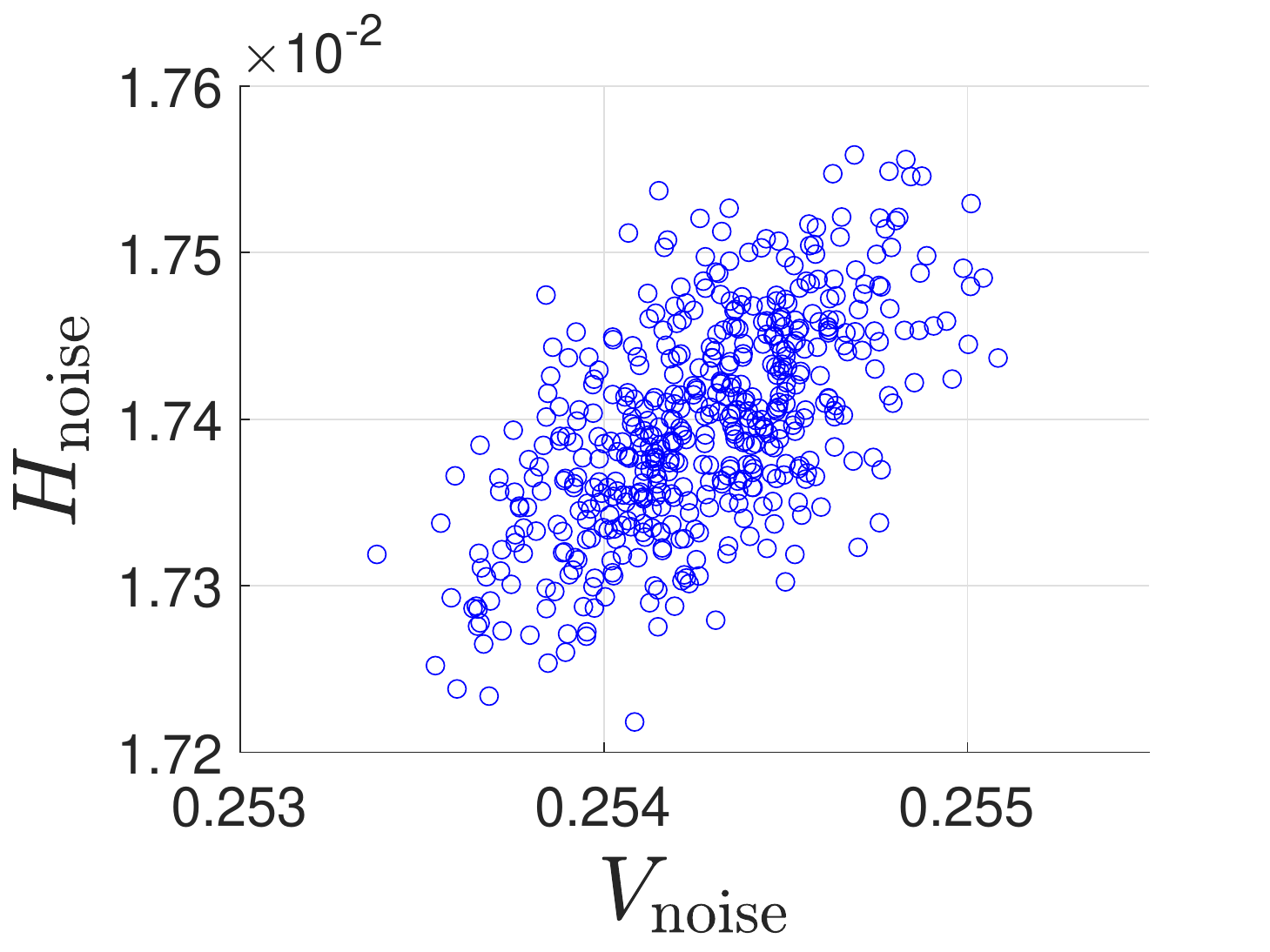}}
    \subfloat[]{\label{fig:3b}\includegraphics[width=.45\textwidth,height=.4\textwidth]{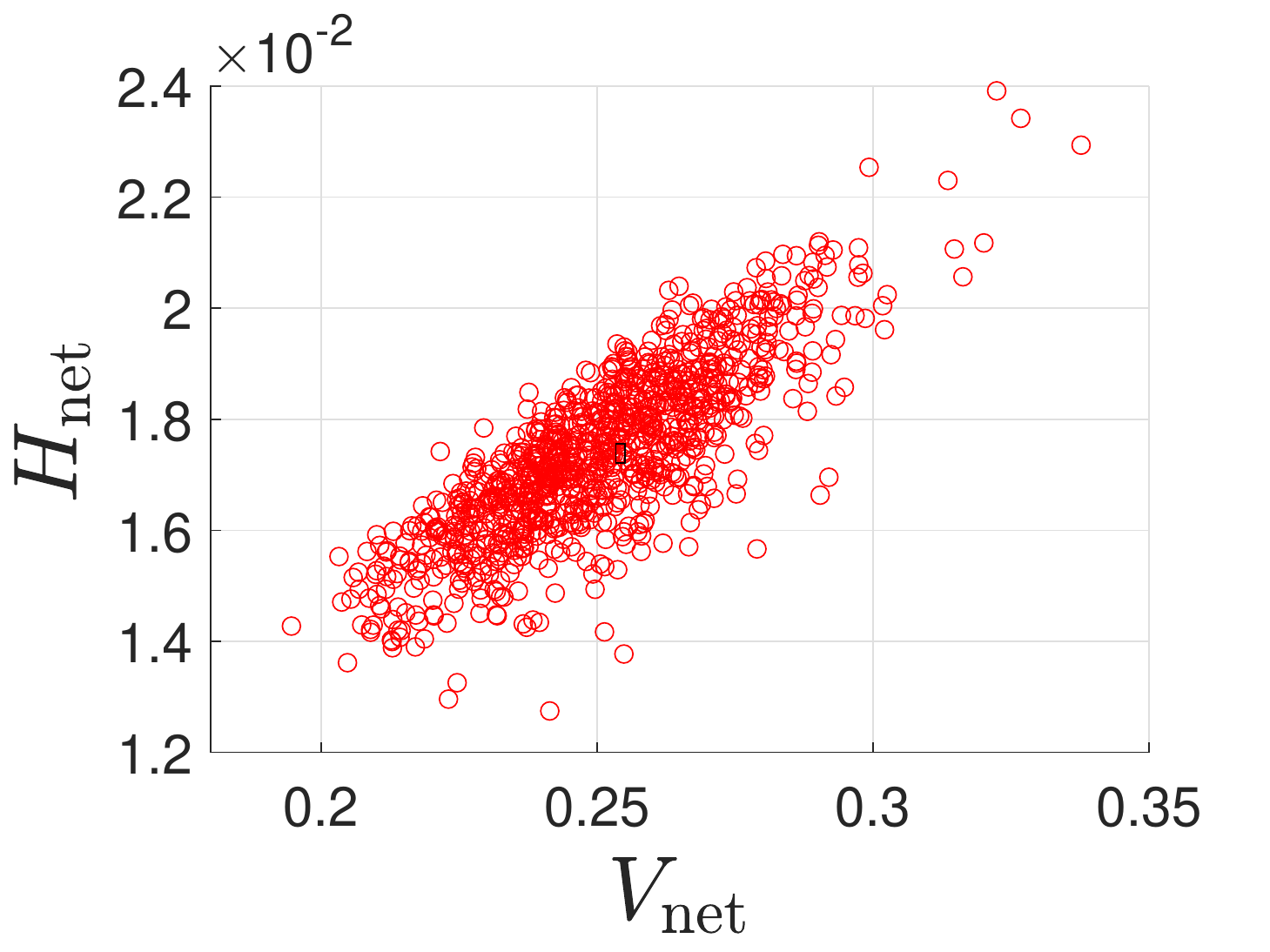}}
    \caption{Scatter plot of throughput versus porosity, under (a) {\color{blue}noise realizations,} (b) {\color{red}network realizations} (with each network perturbed once).  The black rectangle in (b) shows the horizontal and vertical range of (a). For both plots, $\beta = 0.06$.}
    \label{fig:tt_unnormalized}
\end{figure}

\cref{fig:tt_unnormalized} shows throughput versus porosity under (a) noise and (b) network realizations, respectively. Both figures show that throughput is an increasing function of porosity because more porous filters process more filtrate. Note the different scales in the two figures: in \cref{fig:3b} we also plot a black rectangle that represents the total range of \cref{fig:3a} showing that, before we correct for induced porosity changes, network variations incur much more variation in total throughput than pore size variations in this noise/porosity regime. 

\begin{figure}[!h]
    \centering
    \subfloat[]{\label{fig:4a}\includegraphics[width=.45\textwidth,height=.4\textwidth]{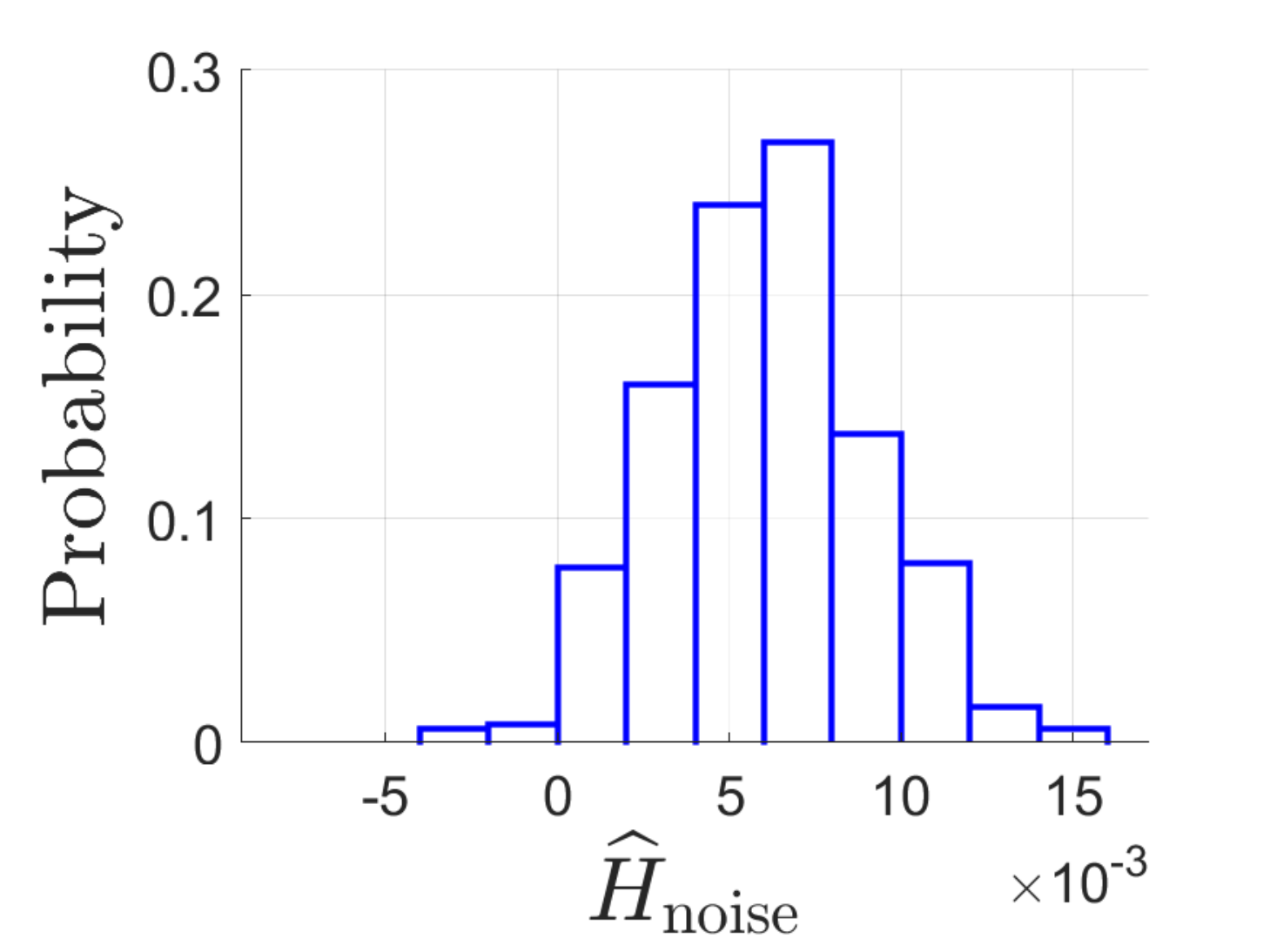}}
    \subfloat[]{\label{fig:4b}\includegraphics[width=.45\textwidth,height=.4\textwidth]{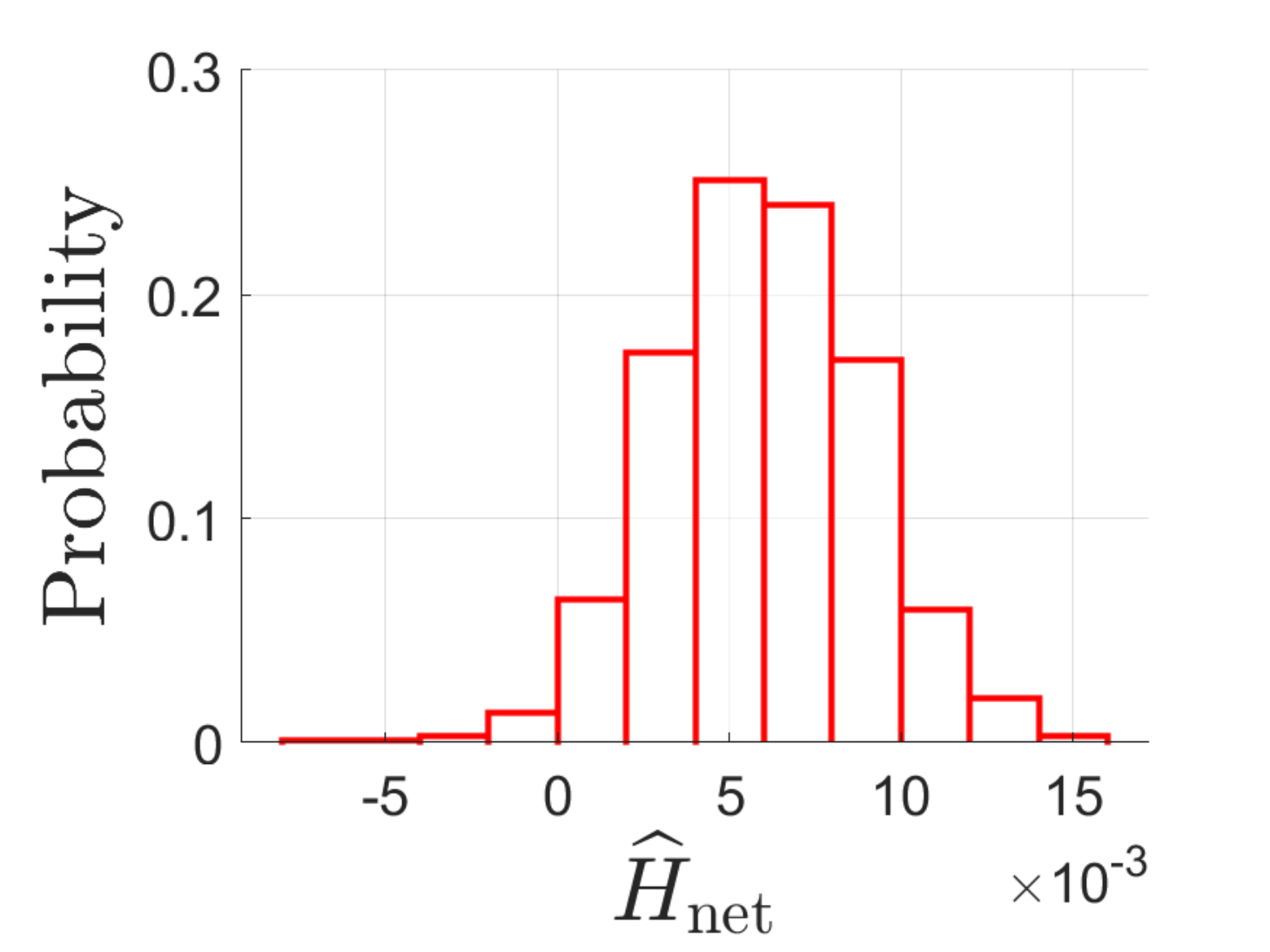}}
    \caption{Histogram of throughput score, under (a) {\color{blue}noise realizations,} (b) {\color{red}network realizations} (with each network perturbed once). Same parameters as in \cref{fig:tt_unnormalized}.}
    \label{fig:tt_score}
\end{figure}

As noted in \cref{sec:methods} (per \cref{eq:perturbation,eq:vol_formula}), and as evident from \cref{fig:tt_unnormalized}, each pore size perturbation leads to a porosity variation. Since initial porosity is known to be a key parameter~\cite{gu_network_2021}, we proceed by considering the {\it noise} and {\it network scores} that correct for induced porosity changes, formulated in \cref{eq:noise_score,eq:network_score} in \cref{algo:step 4} of the algorithm in \cref{sec:methods}. \cref{fig:tt_score} presents histograms of throughput scores under noise and network realizations. We see that the two histograms are very similar. Comparing \cref{fig:tt_unnormalized,fig:tt_score}, we deduce that the porosity change is the crucial factor (at least for the present $(V,\beta)$ values) since \cref{fig:tt_score}, in marked contrast to \cref{fig:tt_unnormalized}, shows that variations of network and noise as measured by the scores have a very similar effect. 

\subsubsection{Concentration and Tortuosity}
We next investigate the influence of noise perturbation on the accumulated concentration of particle impurities in the filtrate (concentration) and the tortuosity of the pore network (see \cref{app:tor} for a detailed definition), quantities that are strongly related in unperturbed pore networks~\citep{gu_network_2021}. 

\begin{figure}[!h]
    \centering
    \subfloat[]{\label{fig:5a}\includegraphics[width=.45\textwidth,height=.4\textwidth]{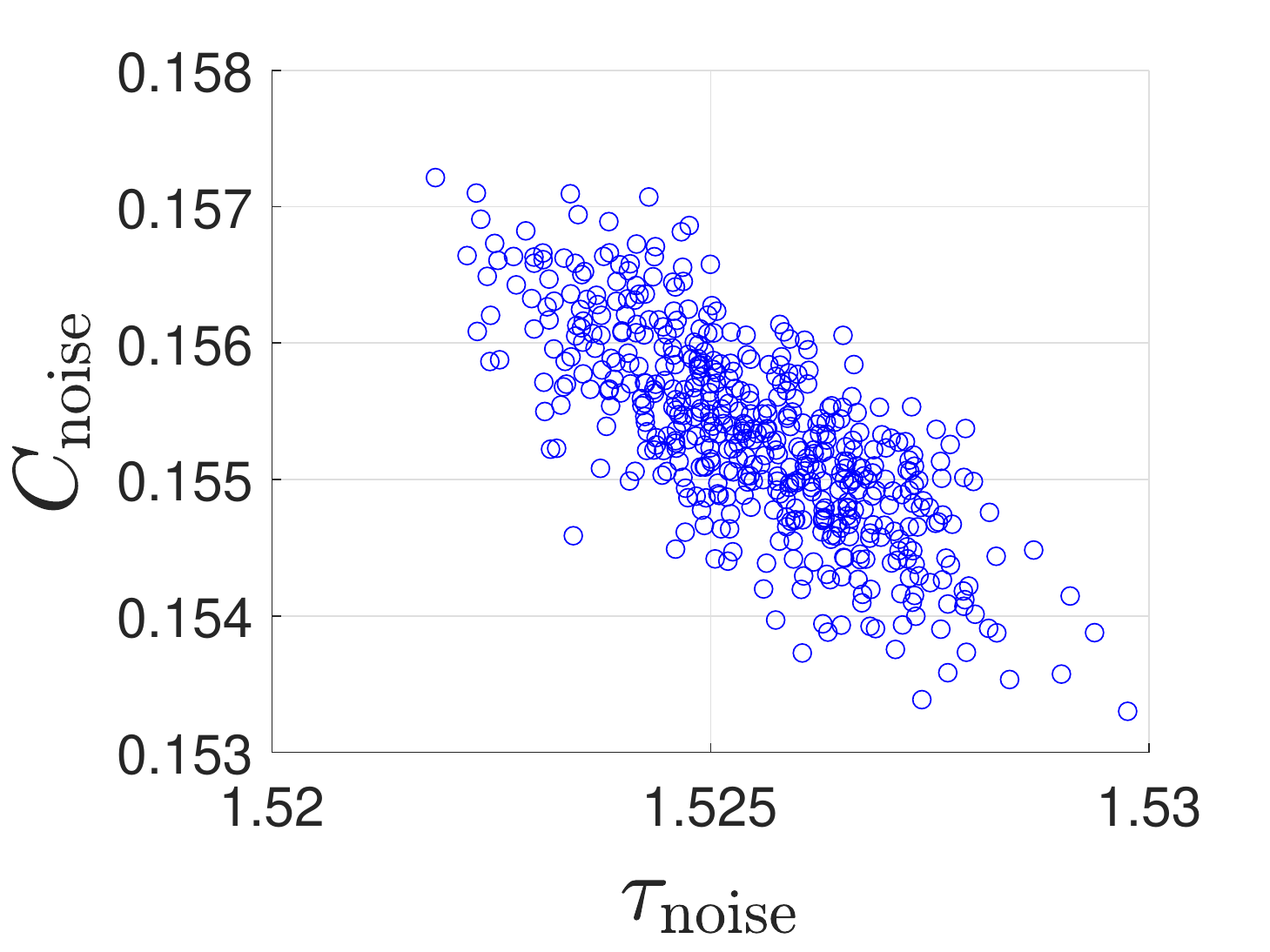}}
    \subfloat[]{\label{fig:5b}\includegraphics[width=.45\textwidth,height=.4\textwidth]{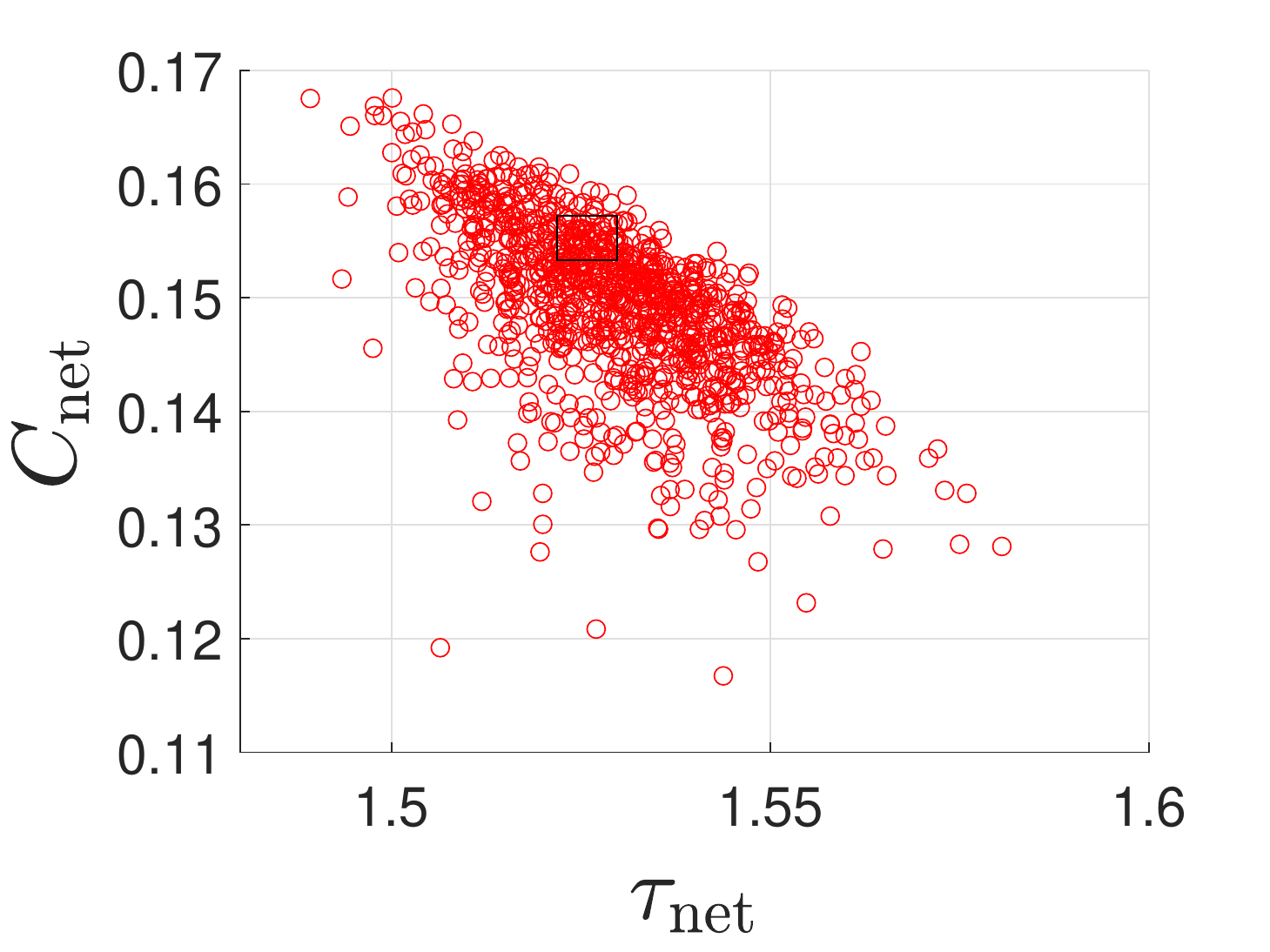}}
    \caption{Scatter plot of concentration versus tortuosity. Same description and parameters as in \cref{fig:tt_unnormalized}.}
    \label{fig:low_poro_conc_tau}
\end{figure}

\cref{fig:low_poro_conc_tau} shows concentration versus tortuosity under (a) noise and (b) network realizations. First, both figures show that concentration is a decreasing function of tortuosity. This makes sense because the longer foulant particles travel in the network, the more likely they are to adsorb to the pore walls, leading to lower concentration at the membrane outlet. Second, we point out that the black rectangle in \cref{fig:5b} covers the total range of \cref{fig:5a}, showing, once again, that before induced porosity changes are accounted for, network variability dominates over noise, for the present choice of parameters.

\begin{figure}[!h]
    \centering
    \subfloat[]{\label{fig:6a}\includegraphics[width=.45\textwidth,height=.4\textwidth]{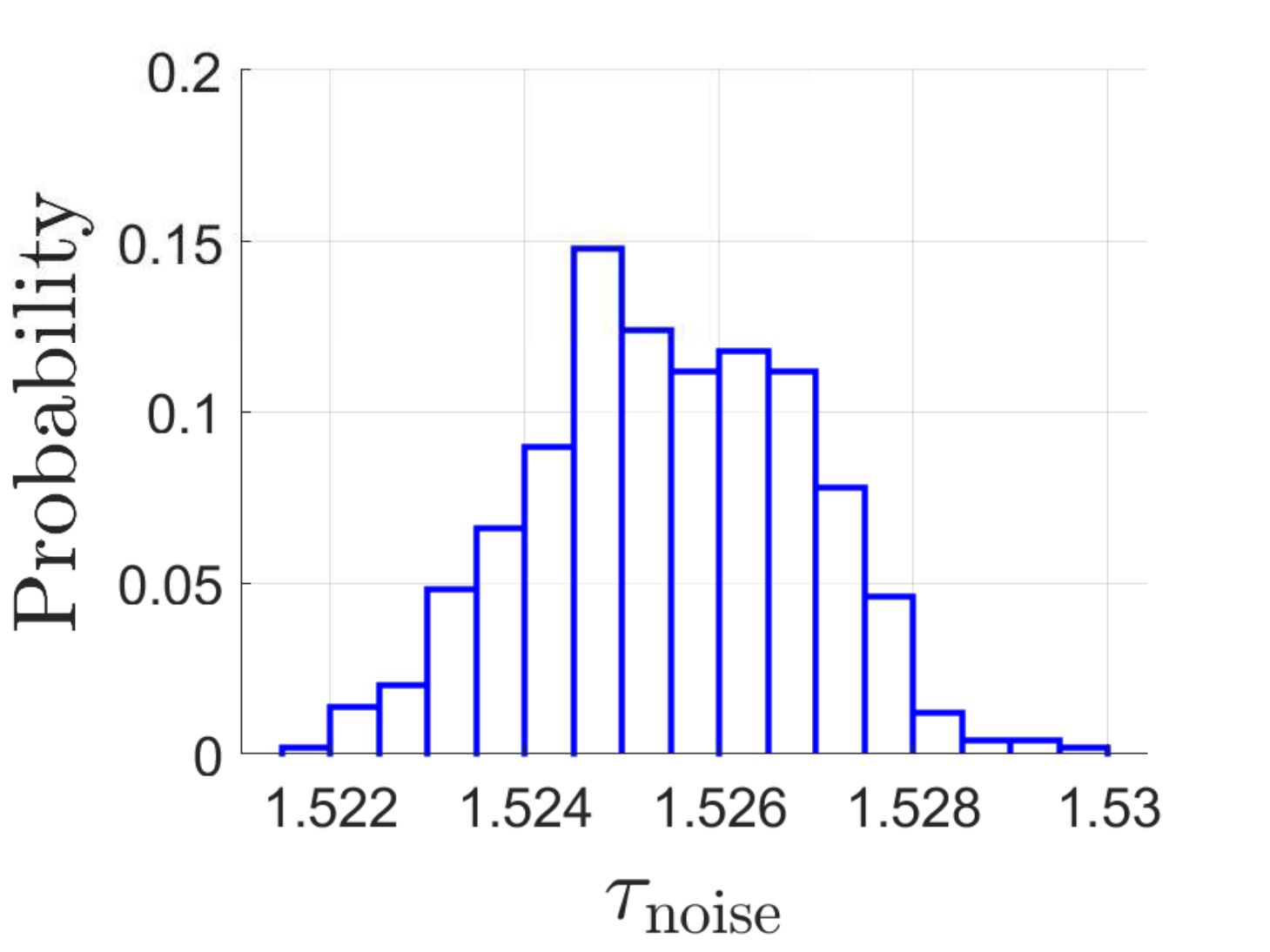}}
    \subfloat[]{\label{fig:6b}\includegraphics[width=.45\textwidth,height=.4\textwidth]{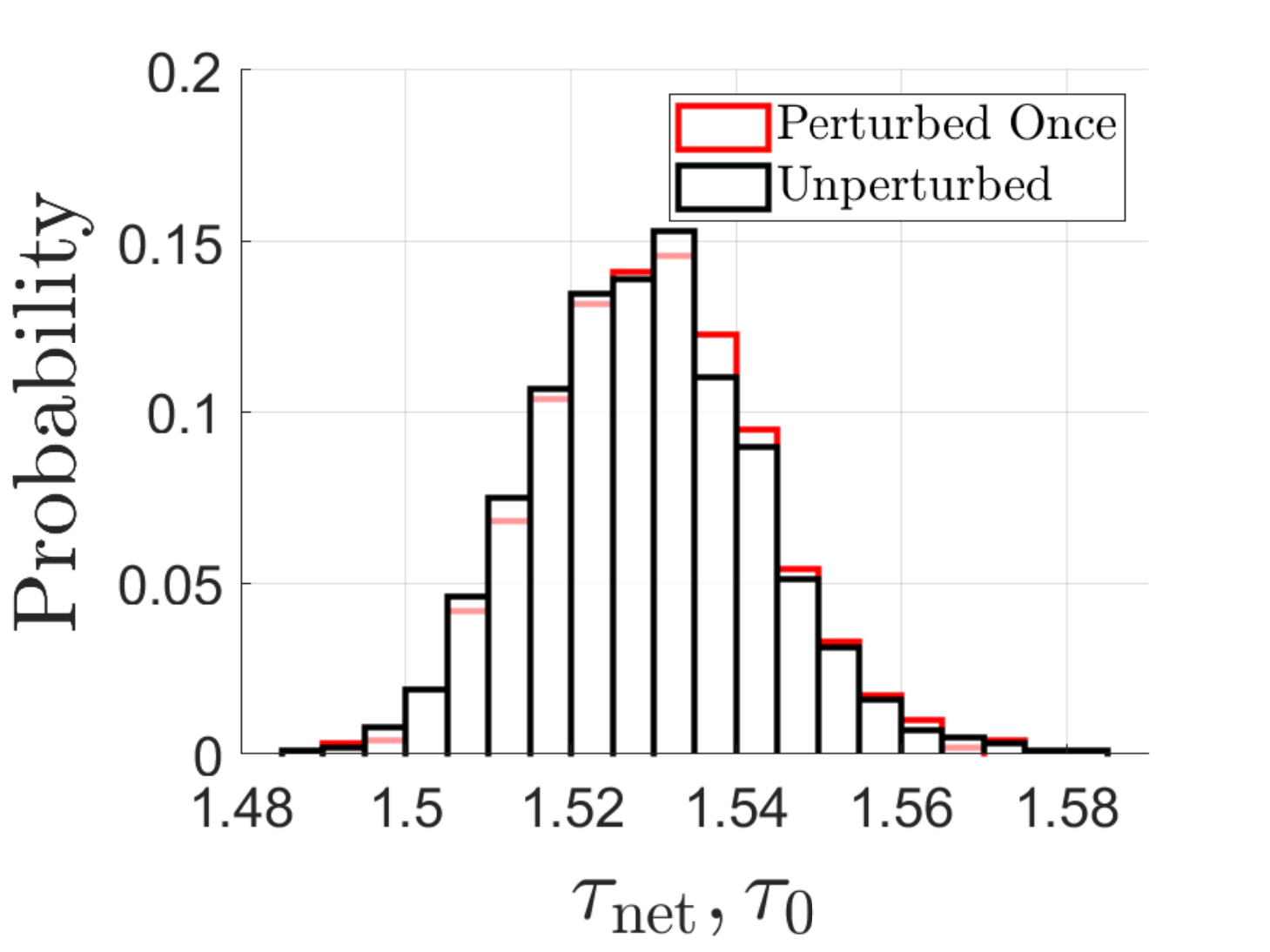}}
    \caption{Histogram of tortuosity under (a) {\color{blue}noise realizations} (b) {\color{red}network realizations}. Same parameters as in \cref{fig:tt_unnormalized}; same data as \cref{fig:low_poro_conc_tau}.}
    \label{fig:low_poro_tau}
\end{figure}

\cref{fig:6a,fig:6b} plot the data of \cref{fig:low_poro_conc_tau} as a histogram of tortuosity under noise and network realizations (respectively). Note the different horizontal ranges on the two plots (cf. \cref{fig:low_poro_conc_tau}). Also, in \cref{fig:6b}, we see that the distribution of tortuosity shifts slightly to the right relative to the unperturbed case, possibly due to some subtle geometric influence from the random network generation, not yet understood.

\begin{figure}[!h]
    \centering
    \subfloat[]{\label{fig:7a}\includegraphics[width=.45\textwidth,height=.4\textwidth]{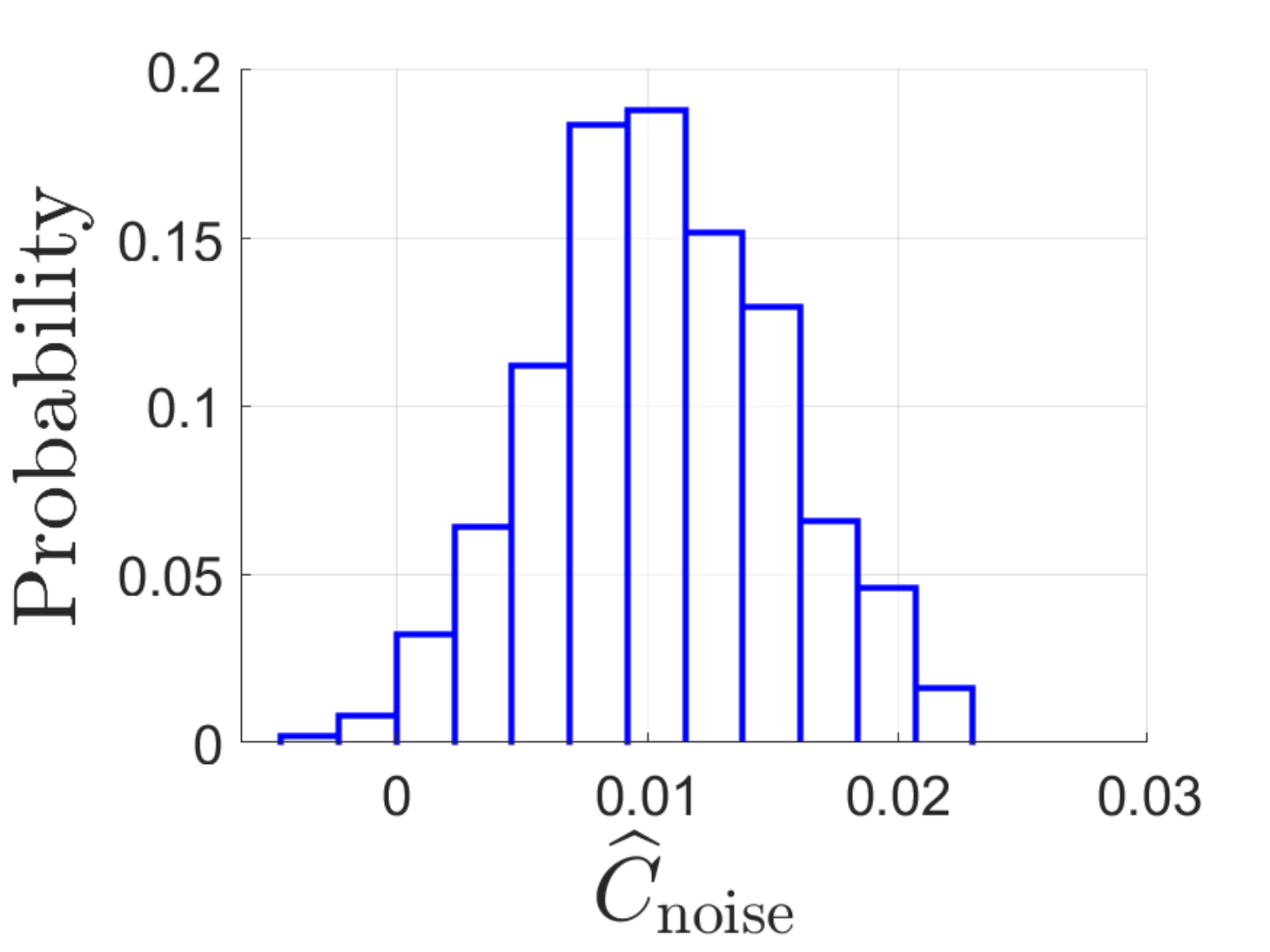}}
    \subfloat[]{\label{fig:7b}\includegraphics[width=.45\textwidth,height=.4\textwidth]{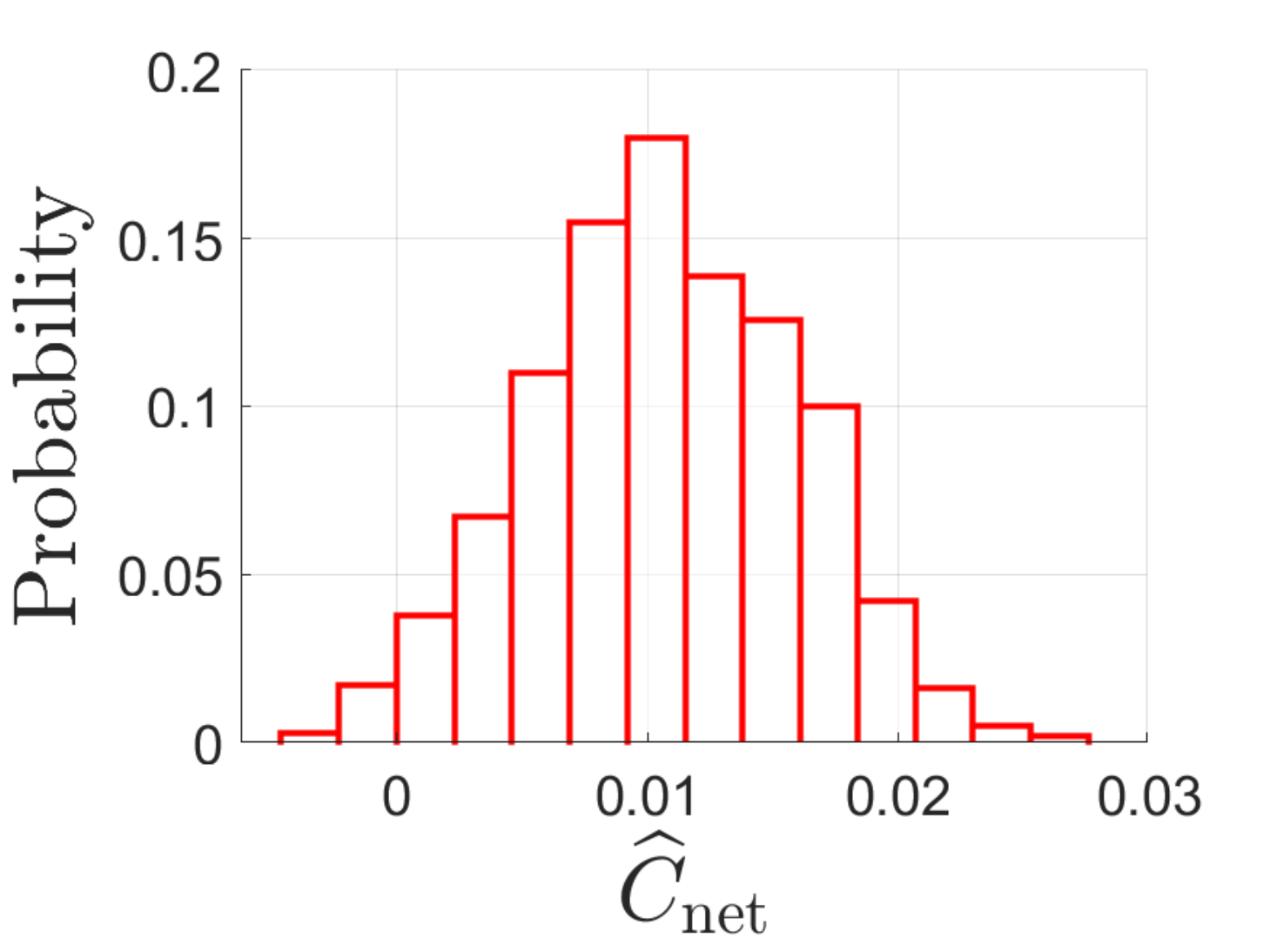}} \\
    \subfloat[]{\label{fig:7c}\includegraphics[width=.45\textwidth,height=.4\textwidth]{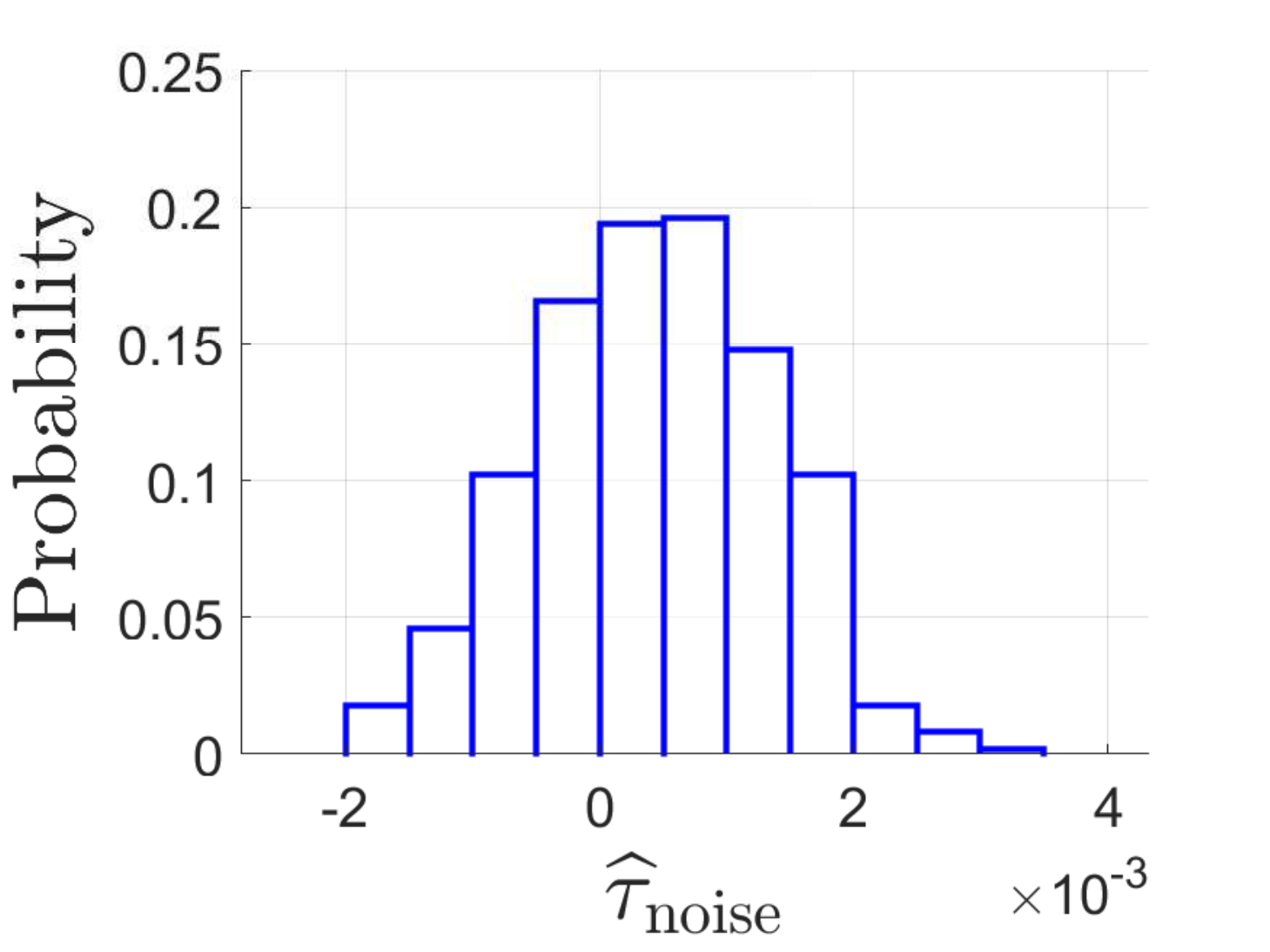}}
    \subfloat[]{\label{fig:7d}\includegraphics[width=.45\textwidth,height=.4\textwidth]{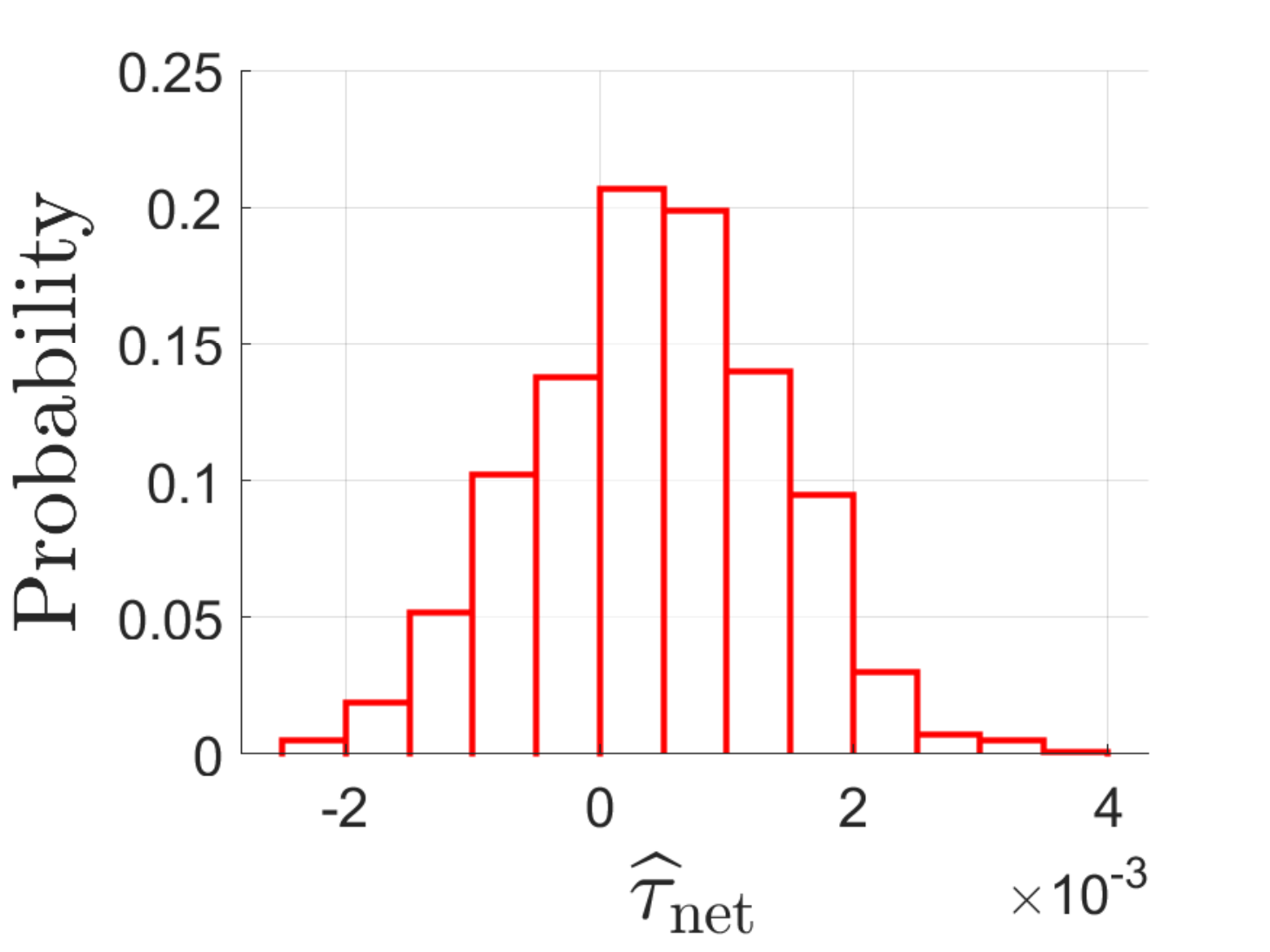}}
    \caption{Histogram of concentration and tortuosity scores, respectively under (a,c) {\color{blue}noise realizations} and (b,d) {\color{red}network realizations}. Same parameters as in \cref{fig:tt_unnormalized}.}
    
    \label{fig:low_poro_conc_stat_stab}
\end{figure}

\cref{fig:low_poro_conc_stat_stab} shows histograms for the porosity-corrected concentration and tortuosity scores under noise (\cref{fig:7a,fig:7c}, generated using \cref{eq:noise_score}) and network (\cref{fig:7b,fig:7d}, generated using \cref{eq:network_score}) realizations.  The histograms of concentration scores in \cref{fig:7a,fig:7b} are very similar in shape and width, suggesting that, after we correct for porosity differences, perturbing one network many times is equivalent to perturbing many networks once. 

\subsection{Results for Varied Noise Amplitude and Porosity}\label{sec:reinforce}

We now explore the influence of noise amplitude, $\beta$, and of initial network porosity, $V$. For brevity, we condense results for different initial porosity regimes in terms of the {\it average} outputs (see \cref{algo:step 5} of the algorithm) and their standard deviations (shown via error bars; see \cref{outputs} for their notations) plotted as functions of $\beta$. 

\begin{figure}[!h]
    \centering
    \subfloat[]{\label{fig:8a}\includegraphics[width=.45\textwidth,height=.4\textwidth]{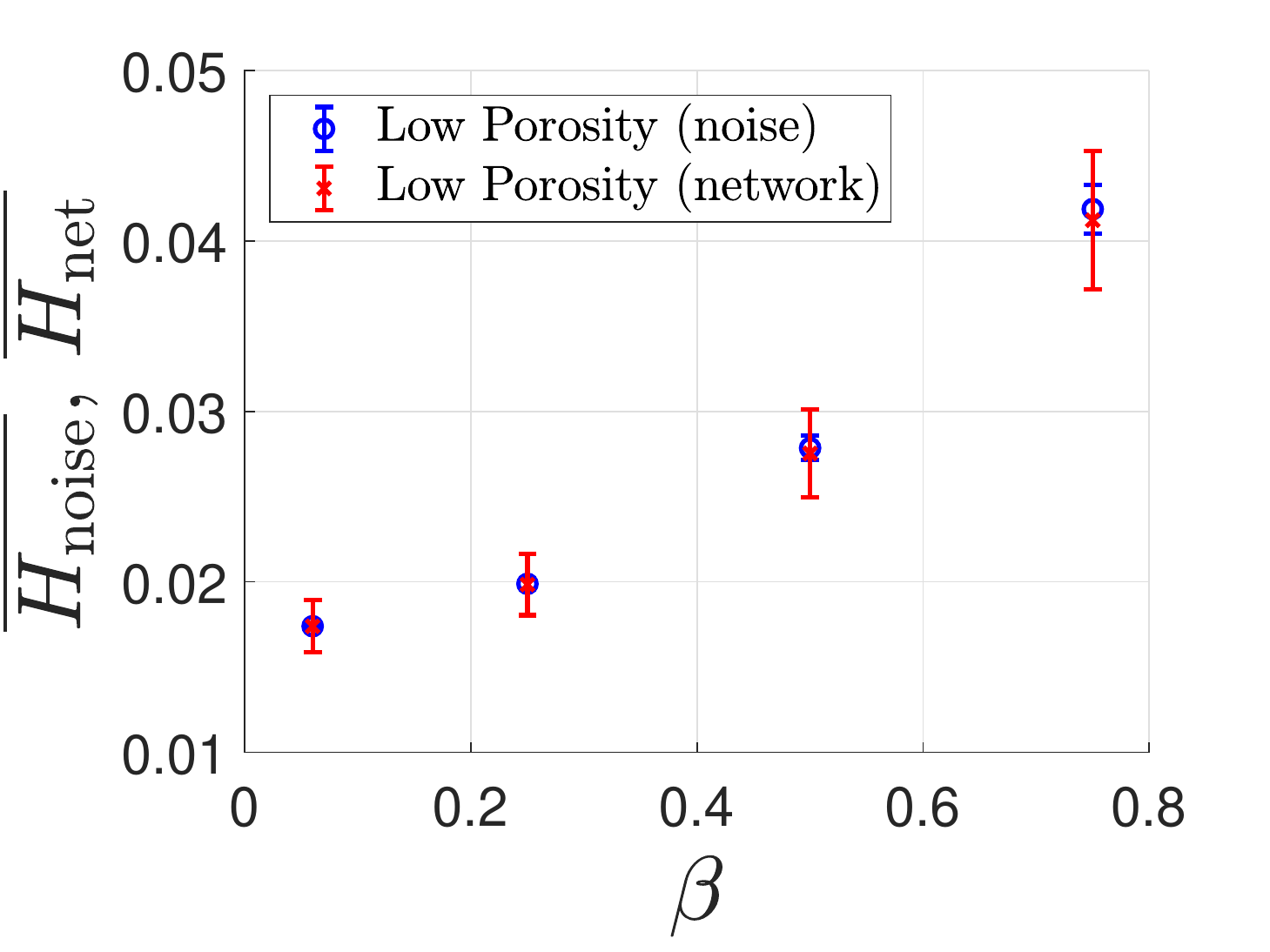}}
    \subfloat[]{\label{fig:8b}\includegraphics[width=.45\textwidth,height=.4\textwidth]{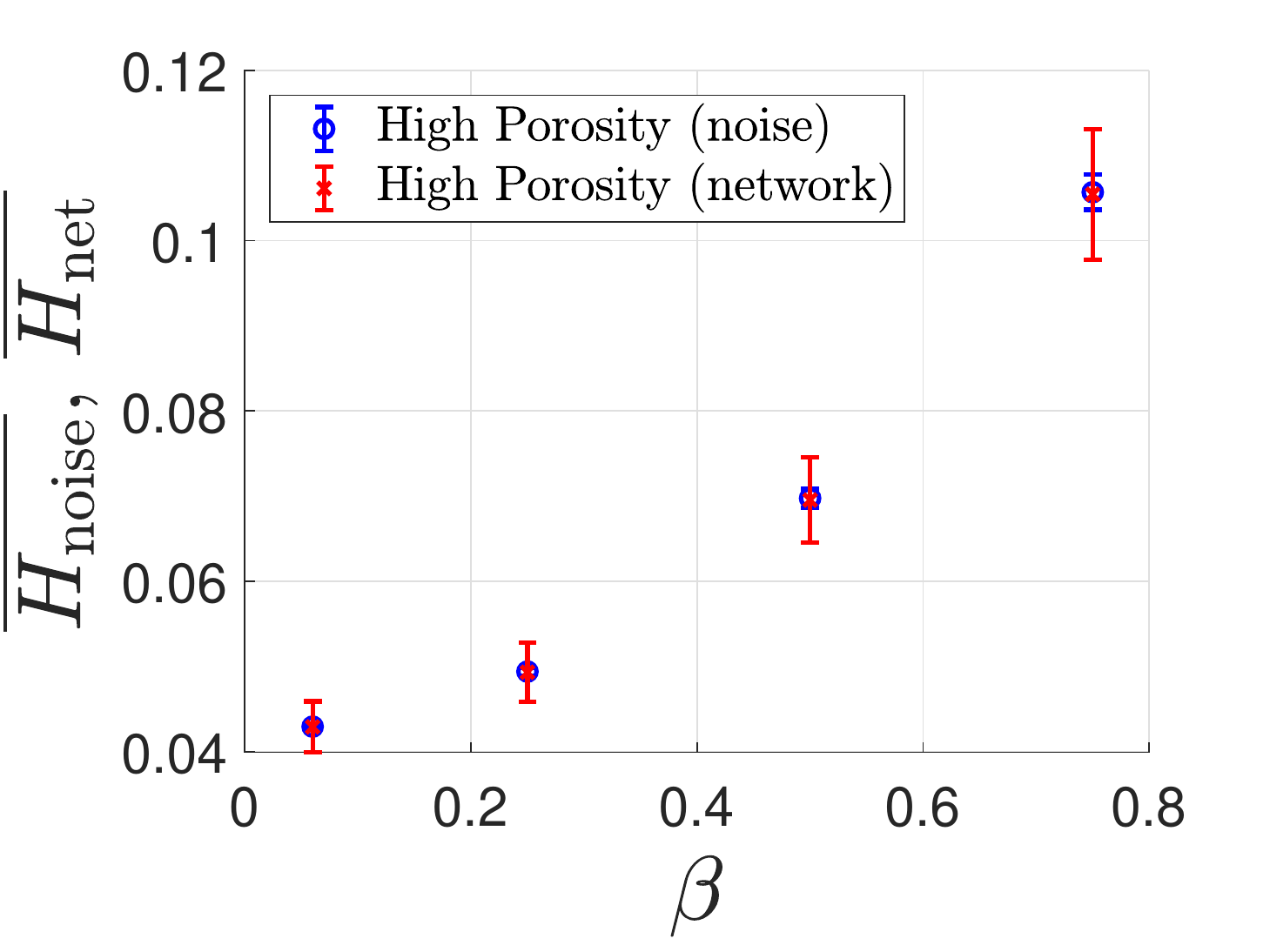}}\\
    \subfloat[]{\label{fig:8c}\includegraphics[width=.45\textwidth,height=.4\textwidth]{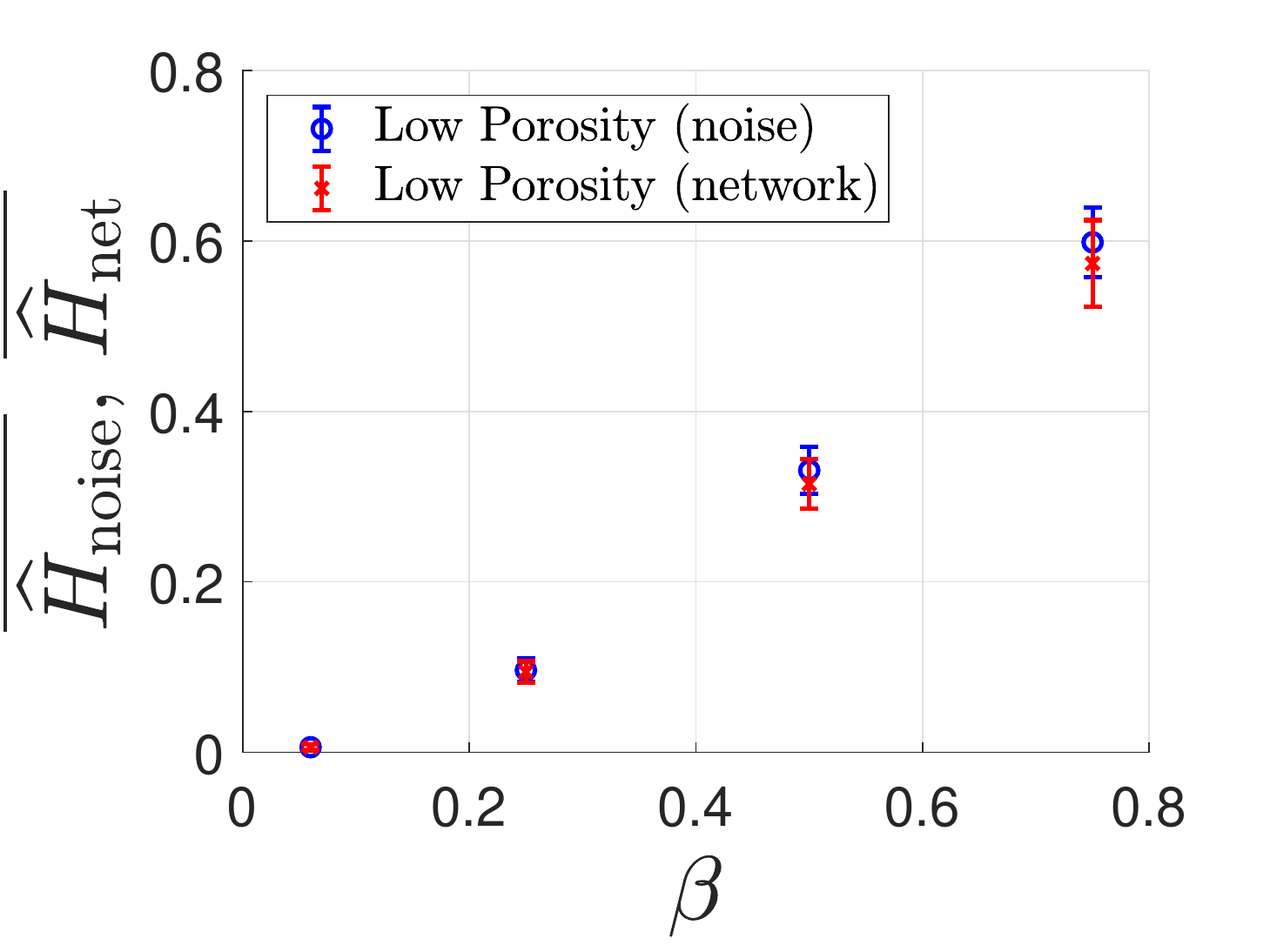}}
    \subfloat[]{\label{fig:8d}\includegraphics[width=.45\textwidth,height=.4\textwidth]{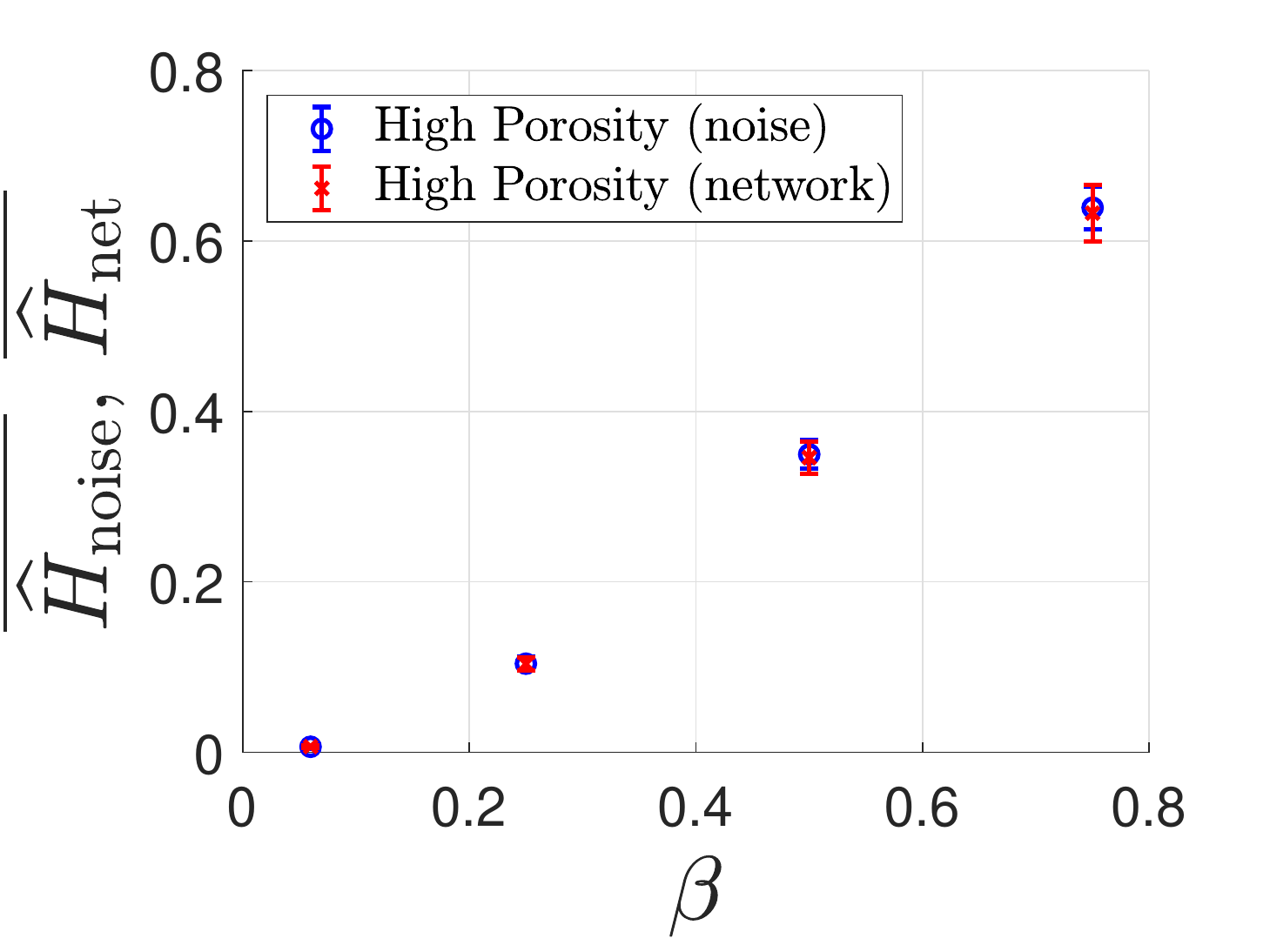}}
    \caption{Mean throughput $\overline{H_{{\rm noise}}},\,\overline{H_{{\rm net}}}$ (a,b) and mean throughput scores $\overline{\widehat{H}_{{\rm noise}}},\,\overline{\widehat{H}_{{\rm net}}}$ (c,d) versus noise amplitude $\beta$ under noise and network realizations. Low porosity $V\approx 0.25$ (a,c) and high porosity $V\approx 0.6$ (b,d) cases are shown. Vertical error bars are standard deviations for each mean value. Note that the scores (c,d) show relative change, so that, e.g., an increase from $0$ to $0.6$ on the vertical axes means a 60\% increase.}
    \label{fig:tt}
\end{figure}

\cref{fig:8a,fig:8b} plot the raw mean throughputs ${\color{blue} \overline{H_{\rm noise}}}$ and ${\color{red} \overline{H_{\rm net}}}$ against noise amplitude $\beta$ for low and high network porosity regimes respectively. These raw mean throughputs are increasing functions of $\beta$, an observation that we attribute largely to the fact that an increase of $\beta$ on average increases initial porosity, which then allows more filtrate to be processed. \cref{fig:8c,fig:8d} plot the mean throughput scores that correct for porosity changes, for filters of low and high porosity, respectively. We find that the mean throughput scores are also increasing functions of $\beta$; however, this increase (which is still significant, around $60\%$ for $\beta = 0.75$ for both low and high porosity networks per \cref{fig:8c,fig:8d}) is due purely to the pore size variations. This implies that, for equal porosity networks, when the magnitude of the noise perturbation $\beta$ is large enough, pore radius variations do lead to an appreciable change in total throughput. This further suggests that pore radius variations promote filtrate production.

\begin{figure}[!h]
    \centering
    \subfloat[]{\label{fig:8_stdlow}\includegraphics[width=.45\textwidth,height=.4\textwidth]{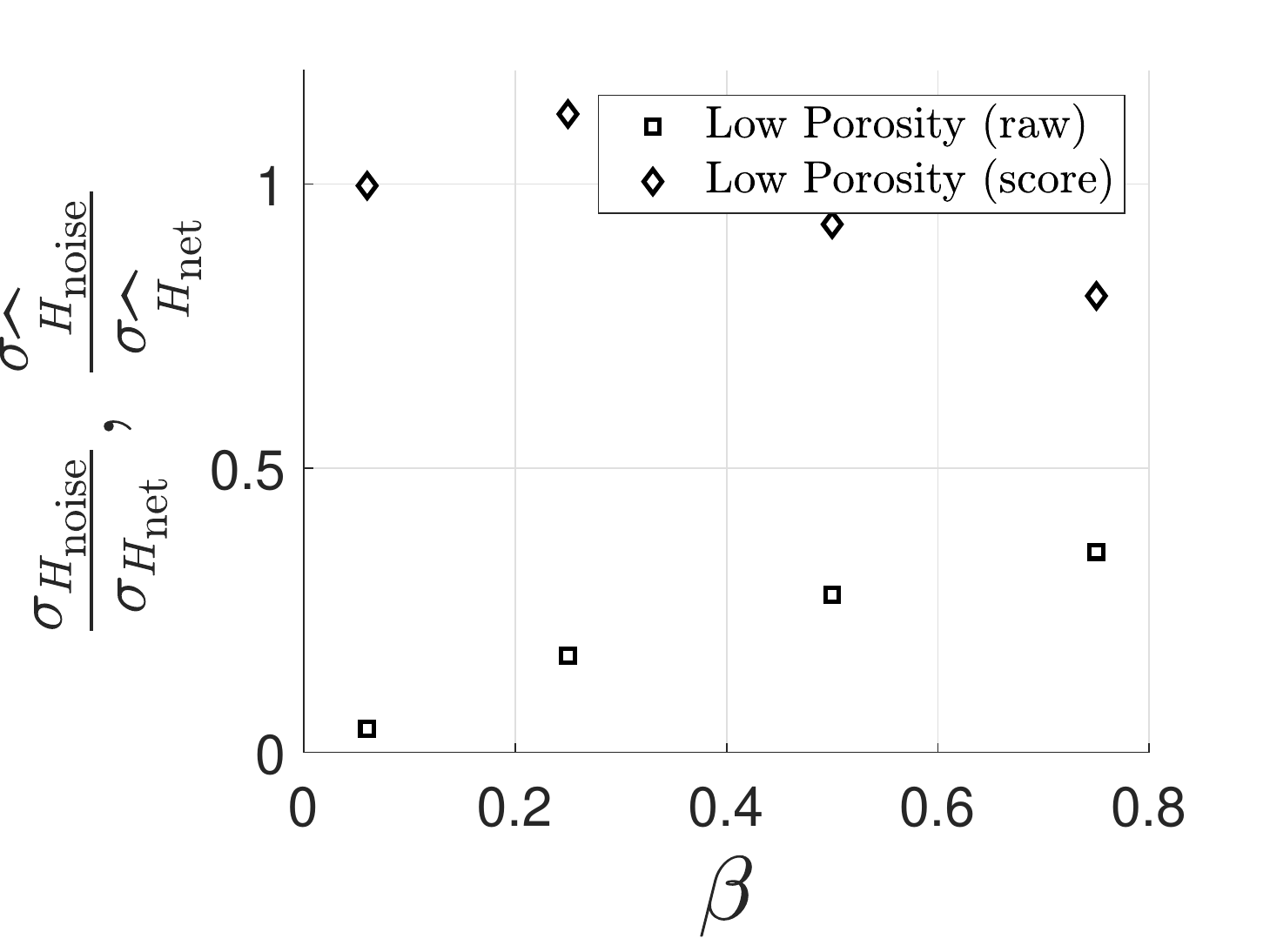}}
    \subfloat[]{\label{fig:8_stdhigh}\includegraphics[width=.45\textwidth,height=.4\textwidth]{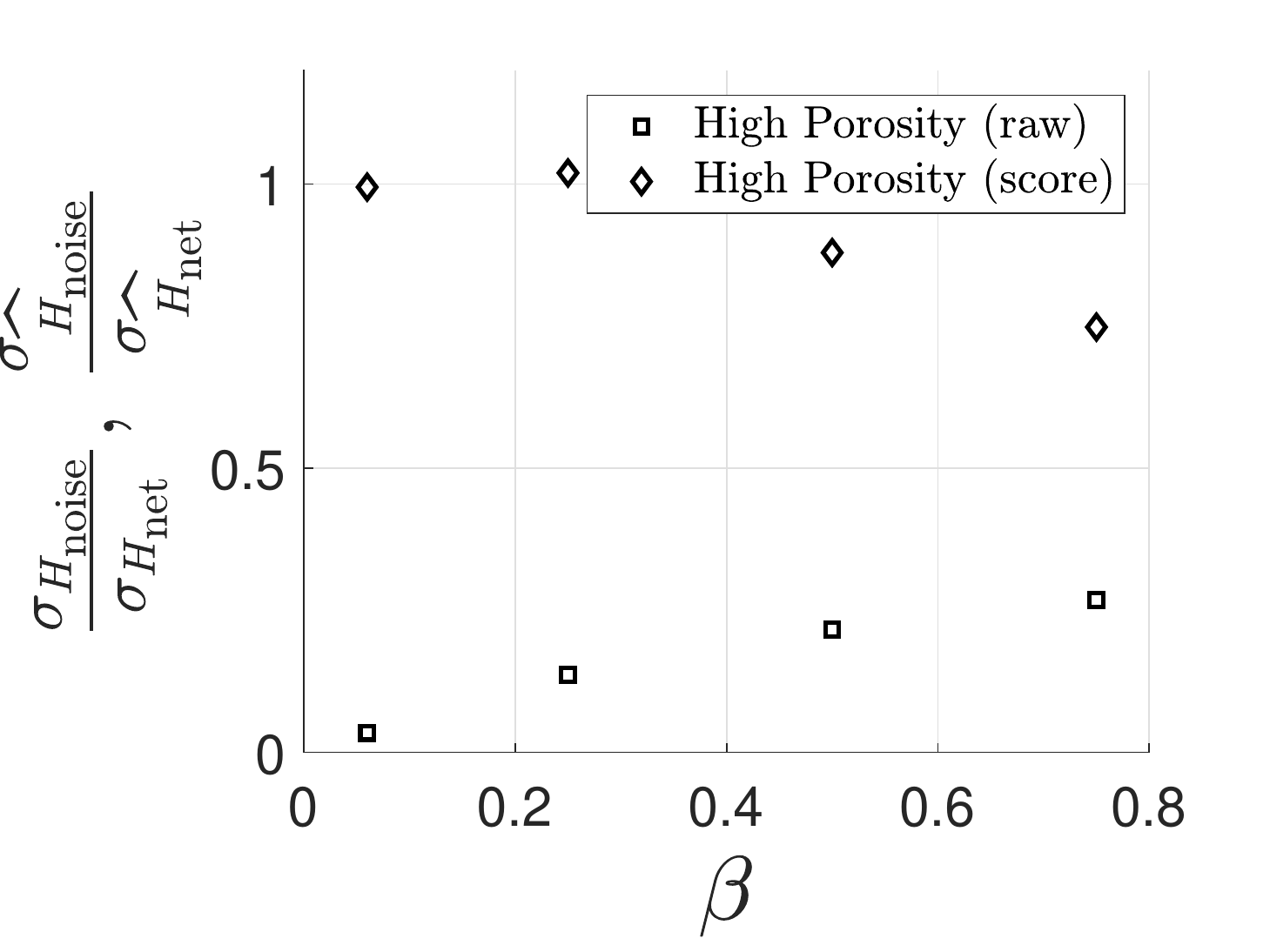}}
    \caption{Ratio of standard deviations of raw throughput (black squares) and throughput scores (black diamonds) for (a) low porosity (corresp. \cref{fig:8a,fig:8c}) and (b) high porosity (corresp. \cref{fig:8b,fig:8d}).}
    \label{fig:tt_std}
\end{figure}

We now study the variability of throughput and throughput scores induced by pore radius variations, represented by the error bars (standard deviation of each statistic) for all cases shown in \cref{fig:tt}.  To facilitate this comparison, we plot in \cref{fig:tt_std} the ratio of the standard deviations of the mean quantities under noise and network realisations (referred to simply as the s.d. ratio below) from \cref{fig:tt}, for raw throughputs and their scores. A value of the s.d. ratio near $1$ implies that the compared quantities have similar standard deviations, whereas a value near $0$ says that variation induced by network realisations dominates that from noise. In both low and high porosity regimes (\cref{fig:8_stdlow,fig:8_stdhigh} respectively), we observe that the s.d. ratios for the raw throughput (black squares) are all below $0.5$ for any value of noise amplitude $\beta$, thus the throughput variability from network realisations dominates that from noise. However, the s.d. ratios of the throughput scores (black diamonds) are close to $1$ in both porosity regimes, for all $\beta$ values considered. This finding shows that network porosity accounts for the differences between network and noise variability.

\begin{figure}[!h]
    \centering
    \subfloat[]{\label{fig:9a}\includegraphics[width=.45\textwidth,height=.4\textwidth]{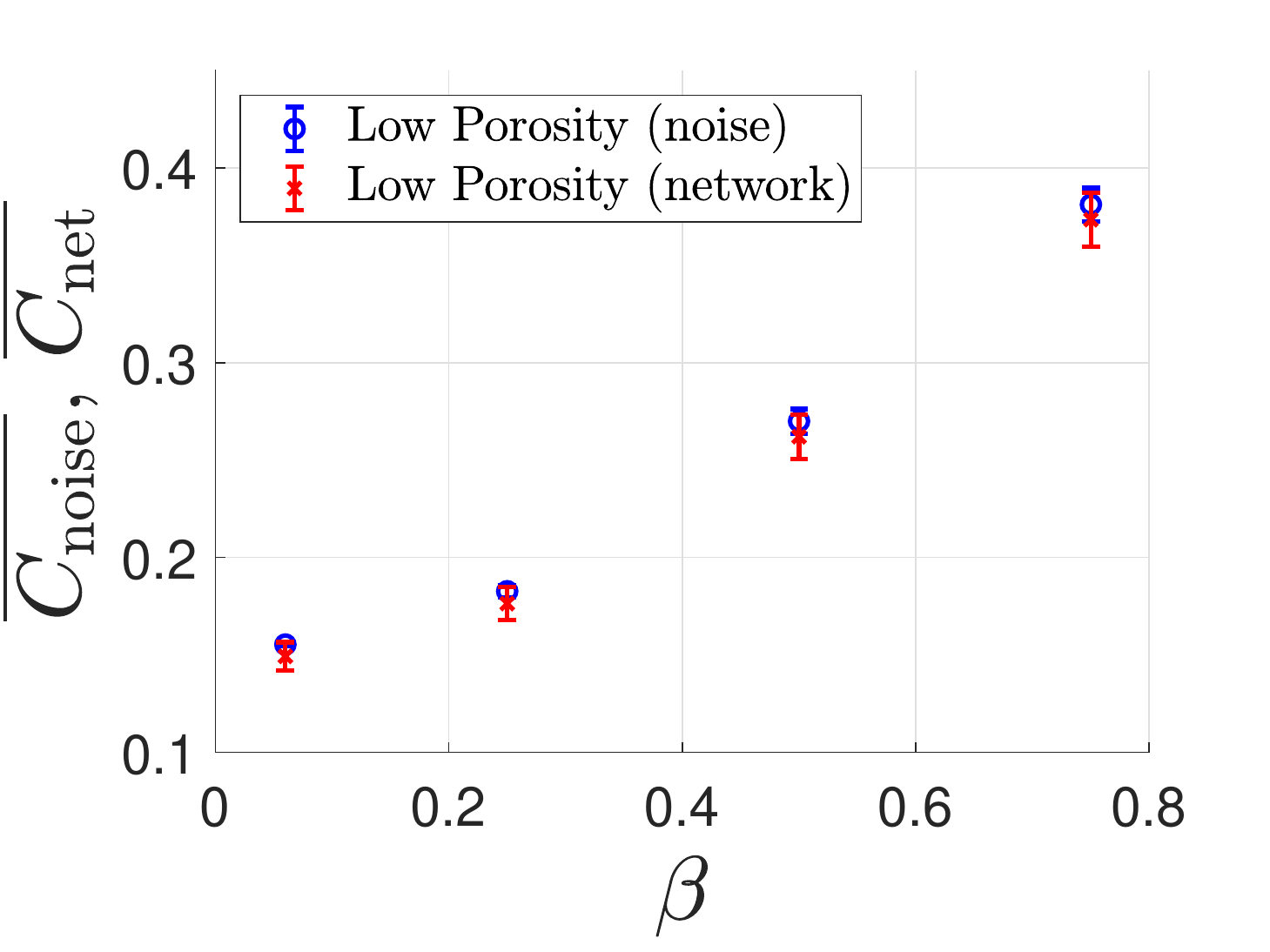}}
    \subfloat[]{\label{fig:9b}\includegraphics[width=.45\textwidth,height=.4\textwidth]{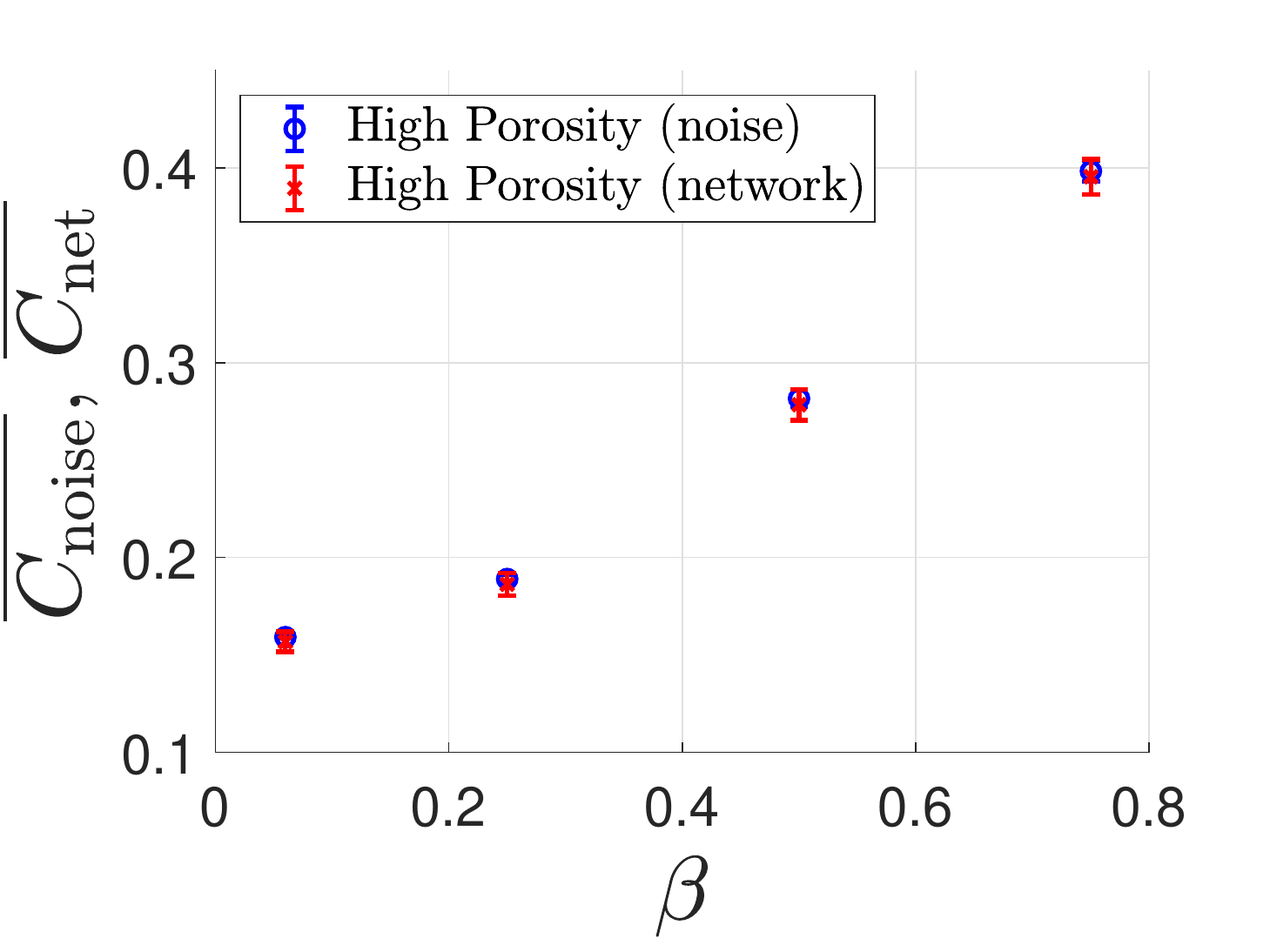}}\\
    \subfloat[]{\label{fig:9c}\includegraphics[width=.45\textwidth,height=.4\textwidth]{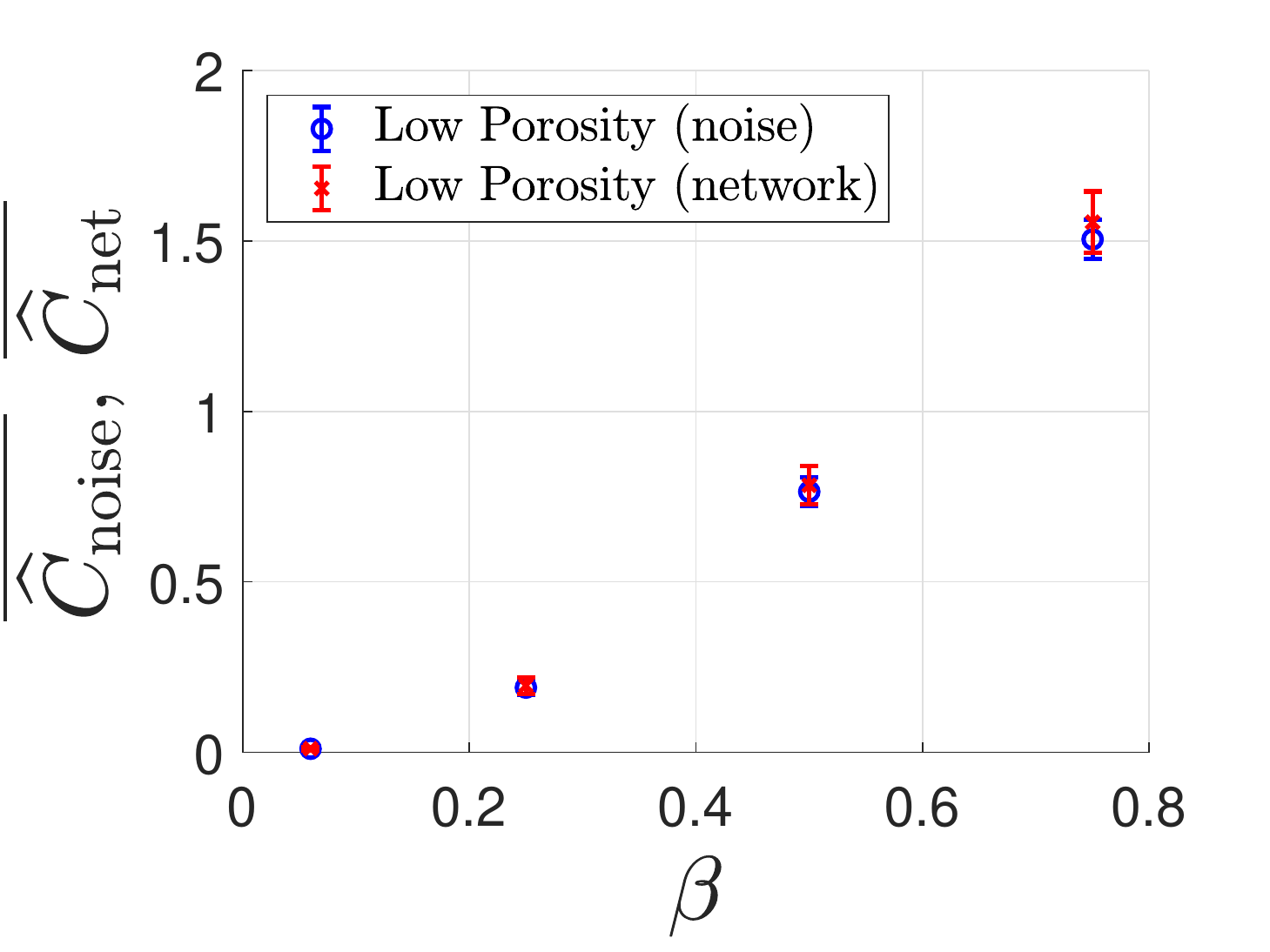}}
    \subfloat[]{\label{fig:9d}\includegraphics[width=.45\textwidth,height=.4\textwidth]{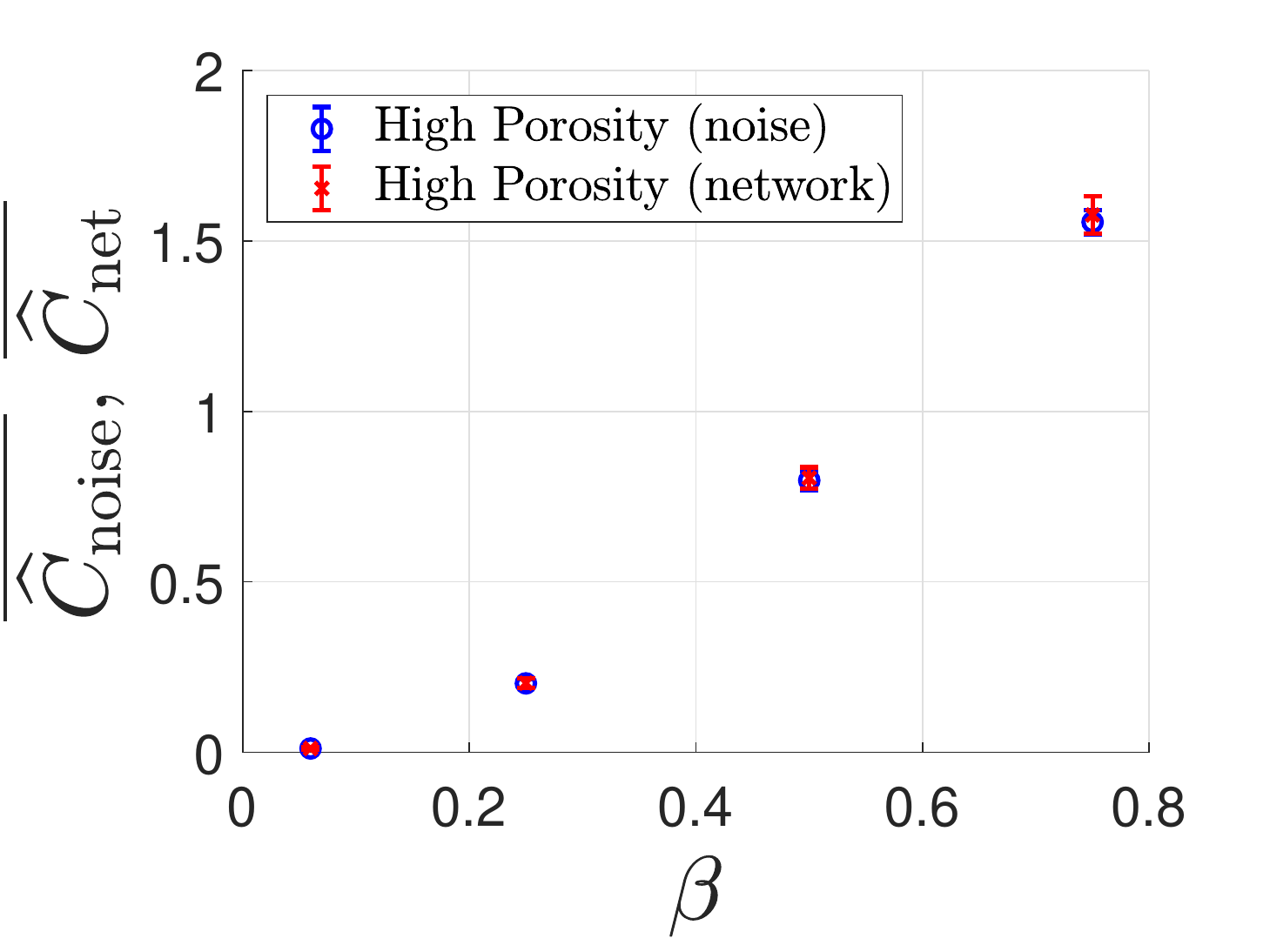}}
    \caption{Mean concentration and concentration score.
    Same setup as in \cref{fig:tt}.}
    \label{fig:conc}
\end{figure}

\cref{fig:9a,fig:9b} show mean concentrations ${\color{blue} \overline{C_{\rm noise}}}$ and ${\color{red} \overline{C_{\rm net}}}$ of particles in the filtrate as a function of noise amplitude $\beta$ for low and high porosity regimes. We observe that the mean raw concentrations in both cases are increasing functions of $\beta$. We attribute this increase largely to the increase of porosity with $\beta$, as evident in~\cref{eq:vol_formula}, allowing more particles to pass through unfiltered. In contrast, \cref{fig:9c,fig:9d} plot the mean concentration scores ${\color{blue} \overline{\widehat{C}_{\rm noise}}}$ and ${\color{red} \overline{\widehat{C}_{\rm net}}}$ (see \cref{eq:noise_score,eq:network_score}), which correct for porosity changes: the increase with $\beta$ seen here is a direct consequence of the pore size variations. For example,
for $\beta = 0.75$, the perturbed networks show more than $150\%$ increase in concentration than the porosity-corrected unperturbed counterparts. This implies that pore radius variations influence foulant concentration in the filtrate significantly for sufficiently large noise amplitude $\beta$, and in fact are detrimental for foulant control.

\begin{figure}[!h]
    \centering
    \subfloat[]{\label{fig:9_stdlow}\includegraphics[width=.45\textwidth,height=.4\textwidth]{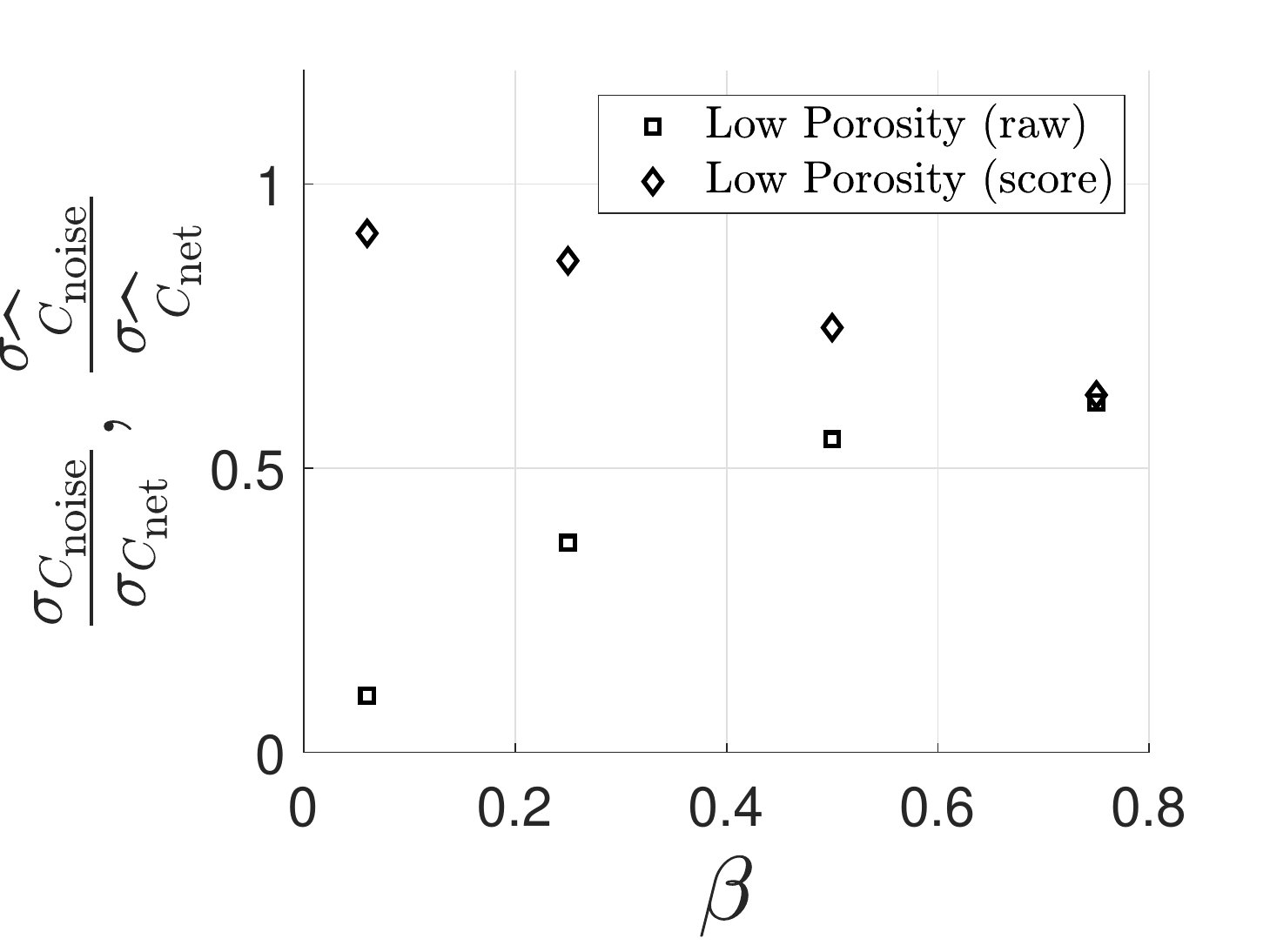}}
    \subfloat[]{\label{fig:9_stdhigh}\includegraphics[width=.45\textwidth,height=.4\textwidth]{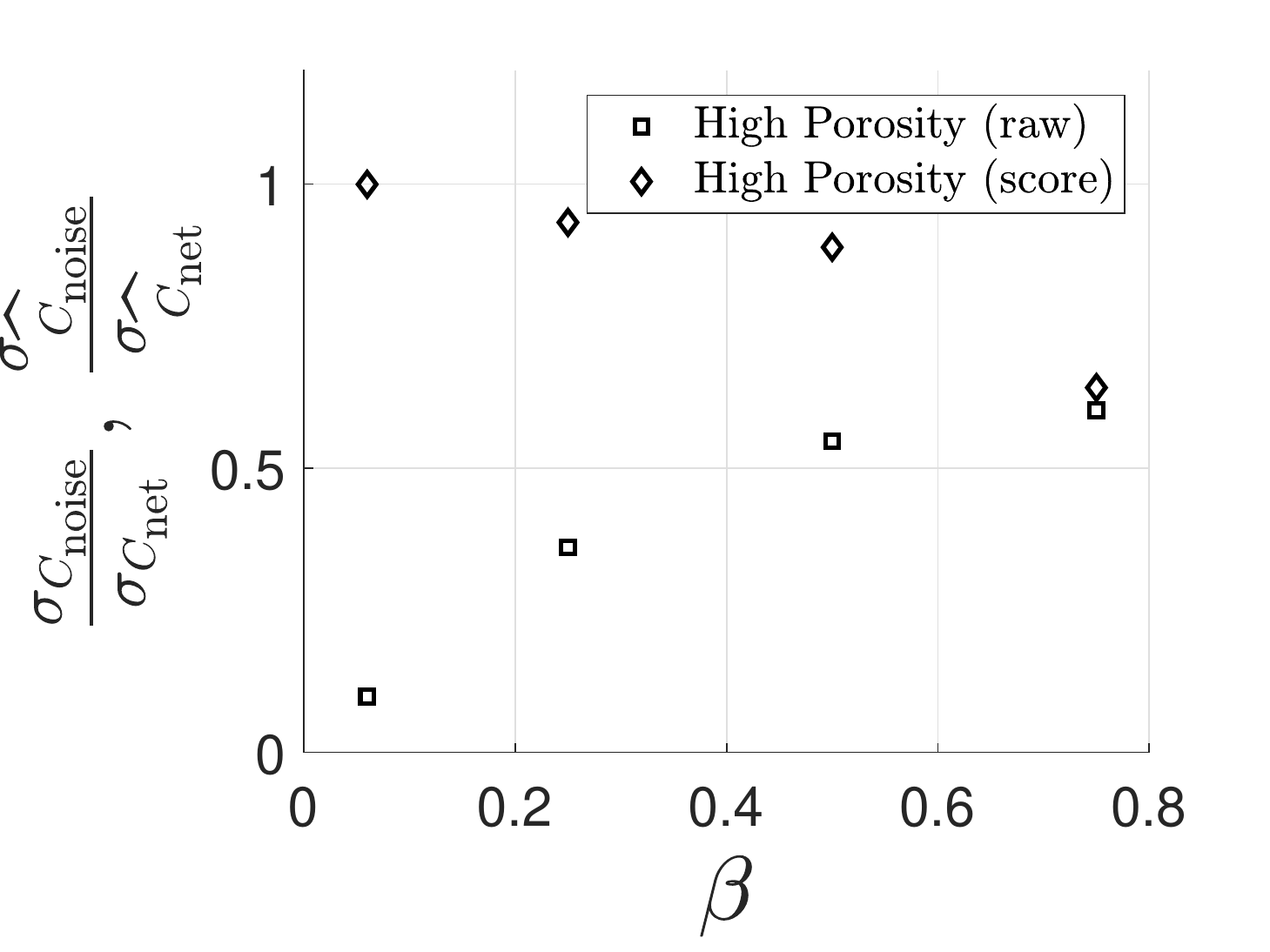}}
    \caption{Same setup as \cref{fig:tt_std} for raw concentration and concentration scores. (a) low porosity (corresp. \cref{fig:9a,fig:9c}) and (b) high porosity (corresp. \cref{fig:9b,fig:9d}).}
    \label{fig:conc_std}
\end{figure}

Now, we discuss the observed variability of the raw mean concentration and concentration scores induced by pore size variations. Similar to the study of throughput variability, \cref{fig:conc_std} plots the ratios of standard deviations of the mean concentrations (raw values and scores) under noise and network realisations, found in \cref{fig:conc}. The observed trends are similar to \cref{fig:8_stdlow,fig:8_stdhigh}: before correcting for induced porosity changes, network variability dominates noise variability for all porosities and noise amplitudes tested, exemplified by the s.d. ratios for the raw concentrations (black squares). However, this dominance is weaker for particle concentration than for throughput: in \cref{fig:conc_std} the s.d. ratios for the raw concentrations are larger than those for the raw throughputs in \cref{fig:tt_std}. This means that concentration, as a performance metric, experiences larger variations from noise perturbation than does throughput. 
We further note that the s.d. ratio of concentration scores (black diamonds) is close to $1$ for small $\beta$ (underscoring the importance of porosity), but decreases as $\beta$ increases. In other words, porosity changes due to network variations are solely responsible for concentration variability at small noise amplitudes, but not at high ones.

\begin{figure}[!h]
    \centering
    \subfloat[]{\label{fig:10a}\includegraphics[width=.45\textwidth,height=.4\textwidth]{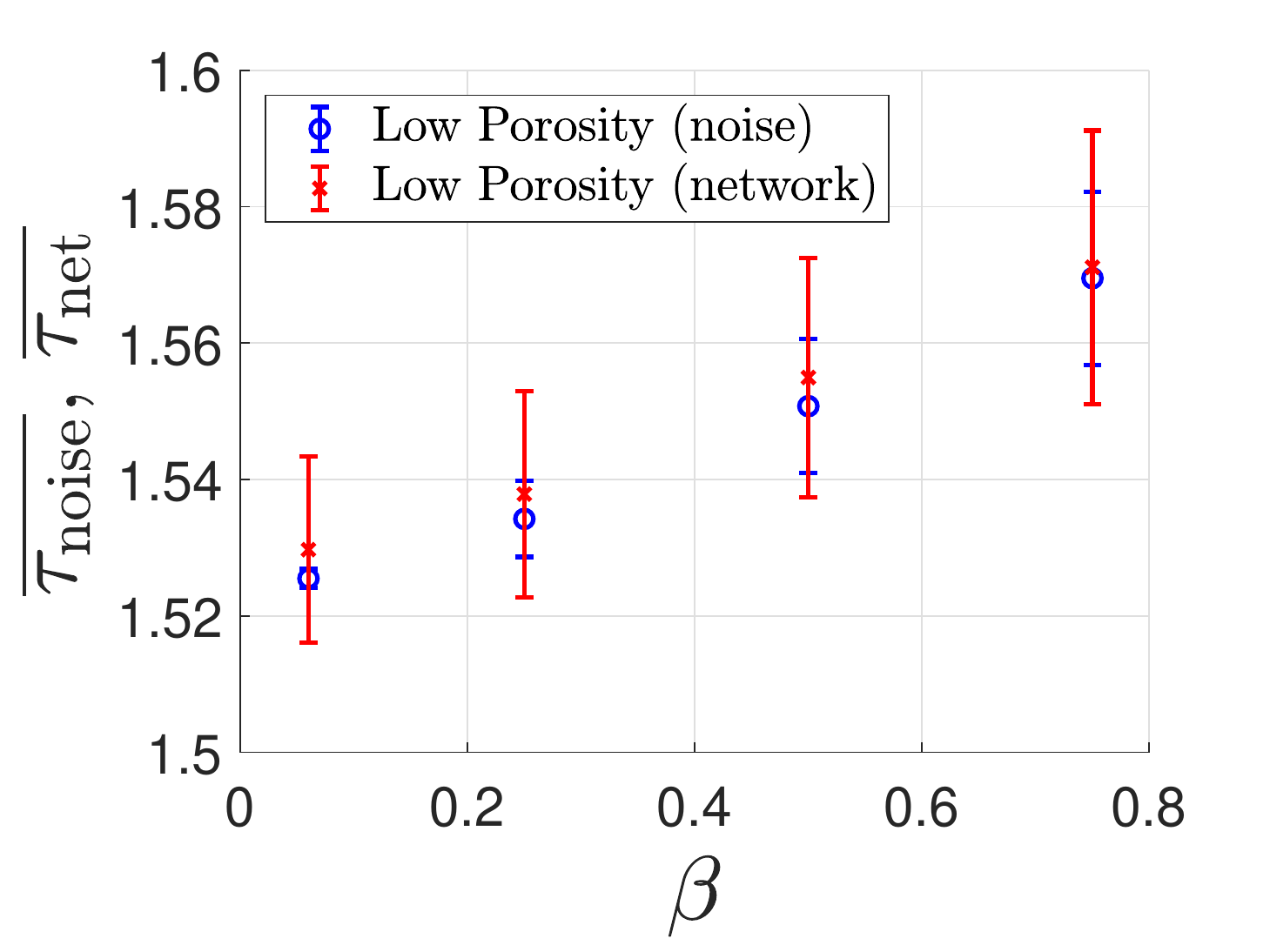}}
    \subfloat[]{\label{fig:10b}\includegraphics[width=.45\textwidth,height=.4\textwidth]{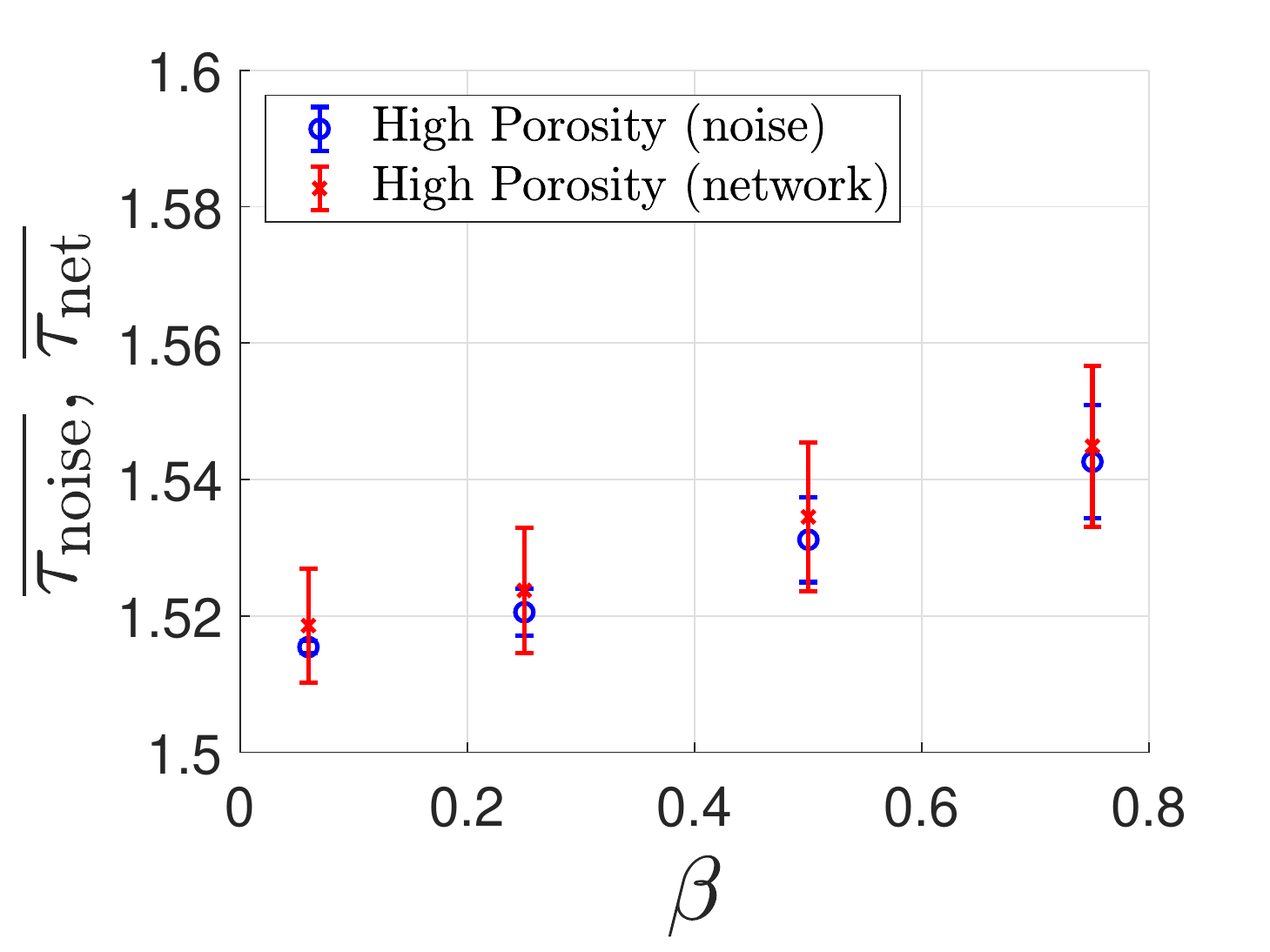}}\\
    \subfloat[]{\label{fig:10c}\includegraphics[width=.45\textwidth,height=.4\textwidth]{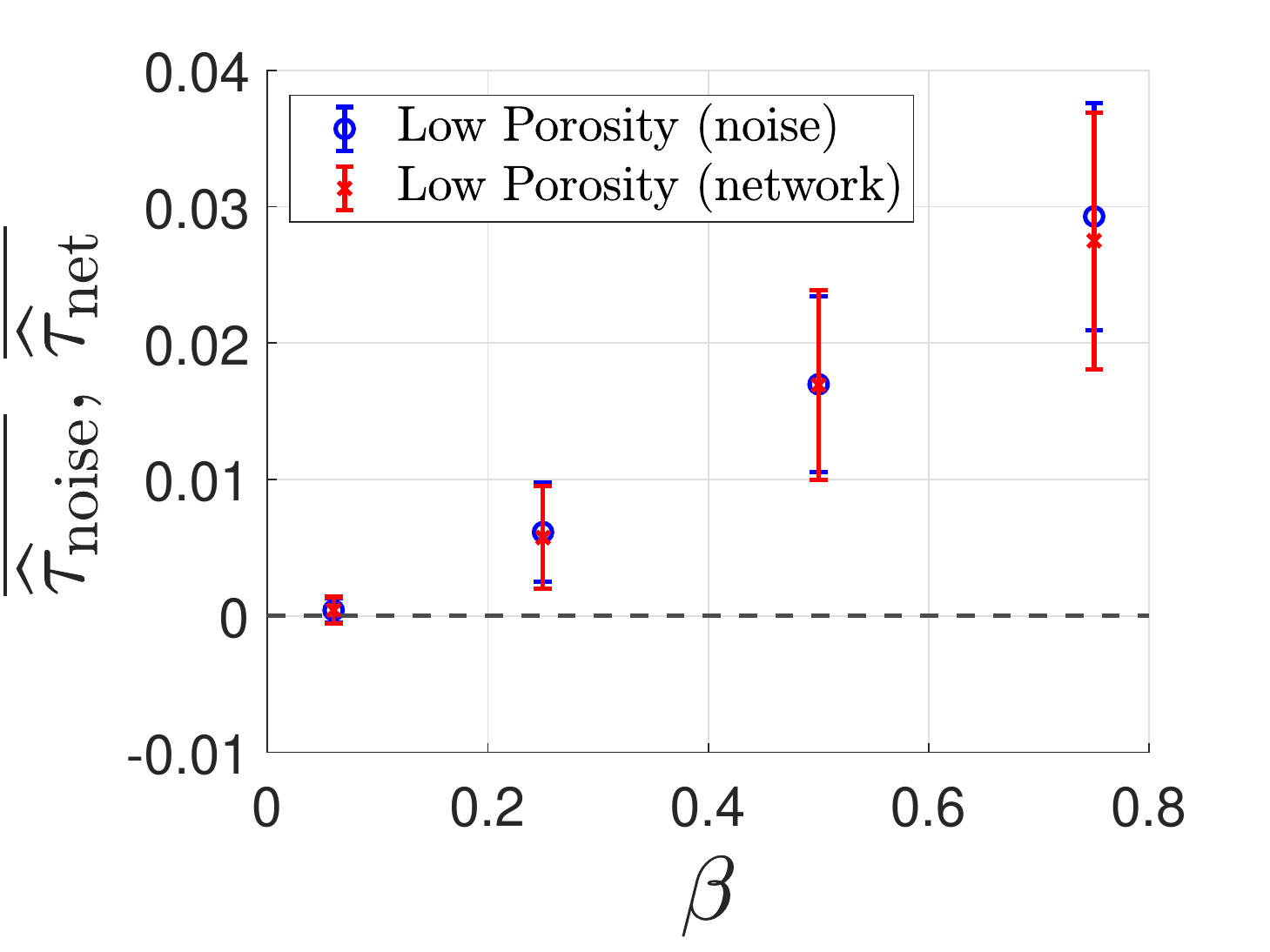}}
    \subfloat[]{\label{fig:10d}\includegraphics[width=.45\textwidth,height=.4\textwidth]{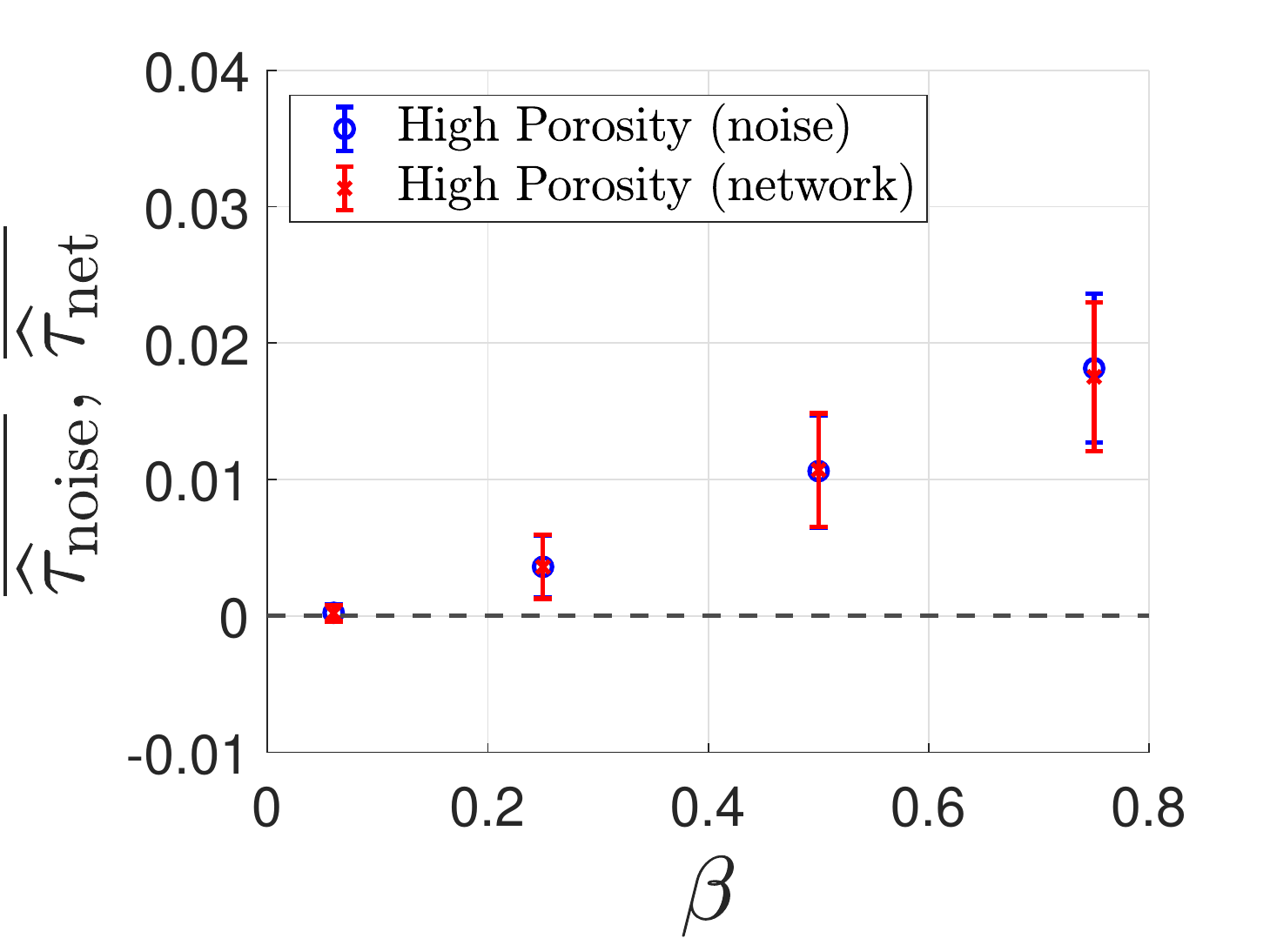}}\\
    \caption{Mean tortuosity and tortuosity score.
    Same setup as in \cref{fig:tt}.}
    \label{fig:tau}
\end{figure}

As noted earlier, tortuosity of an important quantity determining membrane filter performance~\cite{gu_network_2021}.  
\cref{fig:tau} shows mean tortuosities ${\color{blue} \overline{\tau_{\rm noise}}}$ and ${\color{red} \overline{\tau_{\rm net}}}$, and mean tortuosity scores ${\color{blue} \overline{\widehat{\tau}_{\rm noise}}}$ and ${\color{red} \overline{\widehat{\tau}_{\rm net}}}
$ as noise amplitude varies. We find that in both porosity regimes, network variability ({\color{red}red} error bars) still dominates noise variability ({\color{blue}blue} error bars) if we do not correct for induced porosity variations. The dominance is less strong for higher noise amplitudes, however, where the error bar sizes become similar, indicating that sufficiently high noise amplitude induces similar variability in network tortuosity to that due to the network generation protocol. \cref{fig:10c,fig:10d} show the means of the tortuosity scores ${\color{blue}\widehat{\tau}_{\rm noise}}$ and ${\color{red}\widehat{\tau}_{\rm net}}$; we observe that after correcting for porosity changes, noise and network variability are now very similar for all noise amplitudes considered, in both porosity regimes. This implies that, when porosity variations are accounted for, the effects of noise and network variability are comparable. 

Furthermore, we observe that the mean tortuosity and tortuosity scores in \cref{fig:tau} are increasing functions of noise amplitude $\beta$ in all cases, but the changes are modest compared to those in throughput seen in \cref{fig:tt}. Based on this observation, we infer the relative influence of throughput and tortuosity scores on concentration scores. On the one hand, filters with larger throughput scores tend to allow more foulants to pass through and thus produce {\it higher} foulant concentrations in the filtrate. On the other hand, filters with larger tortuosity lead to {\it lower} concentrations~\citep{gu_network_2021}. Hence, as $\beta$ increases, the increased mean throughput and tortuosity scores have opposite effects on mean concentration scores. The fact that we observe still an increased concentration score as $\beta$ increases (per \cref{fig:9c,fig:9d}) implies that the effect of increasing throughput dominates that from increasing tortuosity. 

Before closing, we note in passing some perhaps unexpected features of the results.  
First, comparing \cref{fig:10c,fig:10d}, we observe that filters with higher porosity show a smaller increase in tortuosity with $\beta$. Second, we note that, according to \cref{fig:10c,fig:10d} for the smallest value of $\beta$ used, some networks may have {\it smaller} tortuosity than their unperturbed counterparts, as shown by the first error bar crossing the zero line.

\section{Conclusions}\label{sec:conclusion}
In this work, we have described and implemented a network model for adsorptive fouling of a membrane filter whose pore structure is formed by randomly generating junctions and pores according to a well-defined protocol. Within this network representation, we have modelled pore size variations as random initial conditions (uniform noise with specified amplitude $\beta$) for the pore radii. We have studied these two sources of randomness -- the random network generation protocol and pore radius variations -- by examining and comparing their effects on two key membrane filter performance metrics: total throughput and accumulated concentration of adsorptive foulants. 

Our findings indicate that, unless we account for induced changes in network porosity, the variability in the chosen performance metrics incurred by the random network generation procedure dominates that due to the noise, for noise amplitudes not too large. From this, we conclude that the initial porosity of the pore network is a critical feature of the filter, and a strong determinant of performance.

However, by considering performance scores that account for porosity variations induced by either the random network generation protocol or the random variations of the initial pore radii, we find that the influence of pore radius variations on membrane performance becomes prominent when the noise amplitude becomes large. 

Furthermore, we conclude that pore size variations are favorable for maximizing filtrate production, but unfavorable for foulant control. Although pore size variations do increase pore network tortuosity (favorable for foulant control), this effect is apparently insufficient to lead to improved foulant removal when balanced against the tendency of the increased throughput to transport a greater proportion of foulant particles through the filter. 

We anticipate that including other fouling modes in our model, such as large-particle sieving, will reveal additional aspects of the influence of pore size variations. Considering sieving in addition to adsorption will be a focus of future work. 

\section*{Acknowledgement}
This work was supported by NSF Grants No. DMS-1615719 and DMS-2133255.

\appendix
\section{Appendix}

\subsection{Network Generation}\label{app:gg}
We generate a membrane pore network via a variant of the Random Geometric network. To generate the set of pore junctions $\mathcal{V}$, we place $N_{\rm total}$ randomly distributed points (following a uniform distribution) in a rectangular box of height $2W$, with square cross-section of length $W$. We connect points that lie within a distance of $a_{\rm max}W$, but also at least $\delta W$ apart, where $\delta$ is a control parameter (fixed throughout this work) such that when chosen large enough, it ensures validity of the Hagen-Poiseuille framework used to model fluid flow. These connections form a set of pores $\mathcal{E}$ and together with the junction set $\mathcal{V}$ we obtain an initial network $G = G\left(\mathcal{V},\mathcal{E}\right)$. We say $\left(v_i,v_j\right)\in \mathcal{E}$ when two junctions $v_{i},v_{j}\in \mathcal{V}$ form a pore.

We then cut through the rectangular box with two horizontal planes at heights $0.5W$ and $1.5W$, respectively. The intersections of these two planes and the set of pores $\mathcal{E}$ form the set of inlets $\mathcal{V}_{\rm in}$ and outlets $\mathcal{V}_{\rm out}$ respectively. Altogether, the above procedure forms the domain for fluid flow and fouling, described in \cref{sec:setup}.

\subsection{The Graph Laplacian}\label{app:gl}
We associate each network $G$ with a (weighted) graph Laplacian, a generalization of the finite difference discretization of the classic Laplace operator $\nabla\cdot\nabla$. It is a square matrix whose off-diagonal terms indicate connection weights, and whose diagonal terms record the total weights of neighbors of each discretization point (junction). In our work, the most relevant weight is the conductance $K_{ij}$ of each pore, given by 
\[
K_{ij} = \frac{\pi R_{ij}^4}{8\mu A_{ij}}.
\]
Then the $K$-weighted graph Laplacian is defined as
\begin{equation}
L_{K}:=D-K,
\label{graph_laplacian}
\end{equation}
where
\begin{equation}
D_{ij}=\begin{cases}
\sum_{l=1}^{\left|V\right|}K_{il}, & j=i,\\
0, & \text{otherwise},
\end{cases}
\label{degree}
\end{equation}
where $\left|V\right|$ is the number of junctions.

While the above setup characterizes the flux inside an individual pore, we employ conservation of flux at each interior vertex $v_i$ throughout the network, 
\begin{equation}
    0 = \sum_{v_j: \left(v_i,v_j\right)\in E} Q_{ij}.
    \label{cons_flux}
\end{equation}
Combining \cref{intro_hp,cons_flux}, we form a graph Laplace equation for the pressures $P$ at each vertex, to which we add the specified pressure drop boundary conditions, 
\begin{align}
    L_{K}P\left(v_i\right) &= 0, \quad v_i \in \mathcal{V}_{\rm int}, \\
    P\left(v\right) &= P_0, \quad v\in \mathcal{V}_{\rm in},\\
    P\left(v\right) &= 0, \quad v\in \mathcal{V}_{\rm out}.
    \label{p:bc}
\end{align}
Once the pressures $P\left(v_i\right)$ are found for all interior junctions $v_i \in \mathcal{V}_{\rm int}$, we use \cref{intro_hp} to find flux $Q_{ij}$ in each pore to form a flux matrix $\mathbf{Q}$ with $i$ and $j$ as row and column indices respectively. 

Using conservation of particle flux at each junction, we arrive at the following advection Laplace equation for foulant concentration $C_{i}\left(T\right)$ at each vertex $v_i \in \mathcal{V}_{\rm int}$, 
\begin{align}
L_\mathbf{Q}^{\rm in}C &=\left(\mathbf{Q}\circ B\right)^{\rm T}C_0,\,\, B_{ij} = \exp\left(\frac{-\Lambda R_{ij} A_{ij}}{Q_{ij}}\right),\label{dimless_conc}\\
C_0 &= \left(C_{\rm top},\ldots,C_{\rm top},0,\ldots,0\right)^{\rm T}, \label{dimless_conc_bc}
\end{align}
where $L_\mathbf{Q}^{\rm in}=D_{\mathbf{Q}^{\rm T}}-\left(\mathbf{Q}\circ B\right)^{\rm T}$ is the {\it advection Laplacian} with a sink $B$, whose form arises from an analytical solution to \cref{cont_transport}. ${\rm T}$ and $\circ$ denote matrix transpose and the element-wise multiplication respectively. See \citep{gu_network_2021} for a detailed derivation of this linear system.

\subsection{Performance Metrics}\label{app:pm}
Volumetric throughput of a membrane filter over its operational lifetime is a commonly-used measure of overall efficiency. The volumetric throughput $\widetilde{H}(T)$ through the filter is defined by
\begin{align}
    \widetilde{H}\left(T\right)=\int_{0}^{T}Q_{\rm out}\left(T^{\prime}\right)dT^{\prime}, \\
    Q_{{\rm out}}\left(T\right)=\sum_{v_{j}\in \mathcal{V}_{\rm out}}\sum_{v_{i}:\left(v_{i},v_{j}\right)\in E}{Q}_{ij}\left(T\right),
\end{align}
where $Q_{{\rm out}}\left(T\right)$ is the total flux exiting the filter. With the scales chosen in \cref{eq:scaling}, throughput $\widetilde{H}$ has scale
\begin{equation}
    \widetilde{H} = \frac{W^3}{\alpha C_{0}}H.
    \label{eq:tt_scale}
\end{equation}
In particular, we compute $H_{\rm final}:=H\left(T_{\rm final}\right)$, the total volume of filtrate processed by the filter over its lifetime. 

To connect with experiments and applications, we briefly discuss the order of magnitude of the dimensional throughput $\widetilde{H}$ in \cref{eq:tt_scale}. The parameter $\alpha$ is of the order of foulant (contaminant) particle volume while $C_0$ is the number of particles per fluid (solvent) volume. The product $\alpha C_0$ then yields an estimate of the volume ratio of the (contaminant) solute and the (fluid) solvent. For example, the permissible exposure limit by OSHA (Occupational Safety and Health Administration) for 1-dioxane in contaminated water is at a concentration of about $100$ mg/L \cite{epa}. The density of 1-dioxane is close to that of water, and thus this level of concentration translates to a volume ratio ($\alpha C_0$) of $10^{-4}$. With this estimate and the order of magnitude for dimensionless throughput $H$ seen in \cref{fig:3a} (about $10^{-2}$), $\widetilde{H}$ then is of the order of $10^2 W^3$ where $W^3$ represents the volume of a cubic unit of a membrane filter of side length $W$. This estimate suggests that this cubic unit will process filtrate volume of order $100$ times its membrane material volume, given the parameters used in this work.

Another performance measure is the accumulated foulant concentration in the filtrate, which captures the aggregate particle capture efficiency of the filter.
The accumulated foulant concentration is defined by 
\[
    \widetilde{C}_{{\rm acm}}\left(T\right)=\frac{\int_{0}^{T}C_{{\rm out}}\left(T^{\prime}\right)Q_{{\rm out}}\left(T^{\prime}\right)dT^{\prime}}{\int_{0}^{T}Q_{{\rm out}}\left(T^{\prime}\right)dT^{\prime}},
\]
where 
\[
C_{{\rm out}}\left(T\right)=\frac{{\displaystyle \sum_{v_{j}\in \mathcal{V}_{\rm out}} \sum_{v_{i}:\left(v_{i},v_{j}\right)\in E}}{C}_{j}\left(T\right){Q}_{ij}\left(T\right)}{Q_{{\rm out}}\left(T\right)}.
\]
$\widetilde{C}_{{\rm acm}}$ has scale $\widetilde{C}_{{\rm acm}} = C_{\rm top}C_{\rm acm}$. Of particular interest is $C_{\rm final}:=C_{{\rm acm}}\left(T_{\rm final}\right)$, which provides a measure of the aggregate particle capture efficiency of the filter over its lifetime. 

We further simplify the notations $H_{\rm final}$ and $C_{\rm final}$ to $H$ and $C$ in the main text as we consider only the end states of these performance measures in our analysis.

\subsection{Tortuosity}\label{app:tor}
Tortuosity $\tau$ of a membrane network is defined by the average distance travelled by a fluid particle from membrane top surface to bottom, relative to membrane thickness $W$. We here provide a formula via a probabilistic approach,
\begin{equation}
    \tau=\frac{\pi_{0}^{{\rm T}}}{W}\left(\sum_{n=1}^{m}\mathbf{P}^{n-1}\right){\rm diag}\left(\mathbf{P}\mathbf{W}_{E}\right),
    \label{eq:tor}
\end{equation}
where ${\rm T}$ means vector transpose and ${\rm diag}$ means the diagonal component of a matrix. Here we provide some intuition for each term. The initial distribution $\pi_0$ describes the probability of the fluid particle choosing an inlet on the membrane top surface. To calculate $\pi_0$, we compute the proportion of flux entering each inlet on the upstream surface relative to total flux. $\mathbf{P}$ within the sum describes the law of how a fluid particle traverses the network from one junction to its adjacent neighbors (known as {\it a step}); the upper limit $m$ is the largest number of steps a particle takes to exit the membrane bottom surface, which can be found for each network. Lastly, ${\rm diag}\left(\mathbf{PW}_E\right)$ describes the average distance travelled by the fluid particle in one step starting from each junction. We refer the reader to~\citep{gu_network_2021} for details of the derivation.

\subsection{Porosity Correction}\label{app:vc}
In \cref{algo:step 3} ({\it porosity correction}) of the algorithm, we derive the expression of $r_{\rm pc}$ such that $V_{\rm noise} = V_{\rm pc}$. It relies on writing $V_{\rm pc}$ in terms of $V_0$ (see \cref{algo:step 1} and consider \cref{eq:vol_formula} with $\beta=0$), the porosity of the unperturbed network with initial radius $r_0$:
\begin{align*}
    V_{\rm noise} = V_{\rm pc} &= \frac{\pi}{2} r^2_{\rm pc} \sum_{\text{edge}} \left(\text{edge length}\right) \\
    &= \left(\frac{\pi}{2} r_0^2 \sum_{\text{edge}} \left(\text{edge length}\right)\right) \left(\frac{r_{\rm pc}}{r_0}\right)^2 \\
    &= V_0 \left(\frac{r_{\rm pc}}{r_0}\right)^2
\end{align*}
and thus $r_{\rm pc} = r_0 \sqrt{\frac{V_{\rm noise}}{V_0}}$.

\bibliographystyle{unsrtnat}

\bibliography{arxiv_variable_pore.bib}

\end{document}